\PassOptionsToPackage{table, dvipsnames}{xcolor}
\documentclass[sigconf]{acmart}
\AtBeginDocument{%
  \providecommand\BibTeX{{%
    \normalfont B\kern-0.5em{\scshape i\kern-0.25em b}\kern-0.8em\TeX}}}

\copyrightyear{2024}
\acmYear{2024}
\setcopyright{rightsretained}
\acmConference[CHI '24]{Proceedings of the CHI Conference on Human Factors in Computing Systems}{May 11--16, 2024}{Honolulu, HI, USA}
\acmBooktitle{Proceedings of the CHI Conference on Human Factors in Computing Systems (CHI '24), May 11--16, 2024, Honolulu, HI, USA}
\acmDOI{10.1145/3613904.3641891}
\acmISBN{979-8-4007-0330-0/24/05}

\newcommand{\rr}[1]{{\color{black}{#1}}}

 \newcommand{\xhdr}[1]{\vspace{1.7mm}\noindent{{\bf #1.}}}

\usepackage{comment} % for long comments
\usepackage{xspace}
\newcommand{\githubURL}{\url{https://github.com/behavioral-data/Data-Assistant-Interface}}
\newcommand{\osfURL}{\url{https://osf.io/9x8bj/}}

\newcommand{\woz}{Wizard of Oz\xspace}
\newcommand{\litreview}{literature\xspace}

\newcommand{\inlineheight}{0.5em}
\newcommand{\arrowConsidered}{%
  \begingroup\normalfont
  \includegraphics[height=\inlineheight]{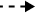}%
  \endgroup
}

\newcommand{\arrowUnaware}{%
  \begingroup\normalfont
  \includegraphics[height=\inlineheight]{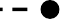}%
  \endgroup
}

\usepackage{soul}
\newcommand{\hlc}[2][yellow]{{%
                  \colorlet{foo}{#1}%
                  \sethlcolor{foo}\hl{#2}}%
}
\newcommand\qt[1]{\hlc[Periwinkle!15]{``#1''}}

\newcommand\shortquote[1]{\qt{\emph{#1}}}
\newcommand\longquote[1]{\qt{\emph{#1}}}
\usepackage{algorithm}
\usepackage{listings}
\usepackage{multirow}
\usepackage{makecell}

\lstdefinestyle{rstyle}{
    language=R,
    basicstyle=\smaller\ttfamily,
    keywordstyle=\color{blue},
    stringstyle=\color{green!60!black},
    commentstyle=\color{gray},
    numbers=none,
    numberstyle=\tiny,
    numbersep=5pt,
    frame=none,
    breaklines=true,
    breakatwhitespace=true,
    showstringspaces=false,
    captionpos=b
}

\begin{document}

%%
%% The "title" command has an optional parameter,
%% allowing the author to define a "short title" to be used in page headers.
% \title{Expanding Analysts’ Thinking: Exploring the Potential of AI-Assisted Data Science through Findings from a Wizard of Oz Study}
\title[How Do Data Analysts Respond to AI Assistance?]{How Do Data Analysts Respond to AI Assistance? A Wizard-of-Oz Study}

%%
%% The "author" command and its associated commands are used to define
%% the authors and their affiliations.
%% Of note is the shared affiliation of the first two authors, and the
%% "authornote" and "authornotemark" commands
%% used to denote shared contribution to the research.

\author{Ken Gu}
\email{kenqgu@cs.washington.edu}
\orcid{0000-0002-4343-1578}
\affiliation{%
  \institution{University of Washington}
  \city{Seattle}
  % \state{Washington}
  \country{USA}
}

\author{Madeleine Grunde-McLaughlin}
% \authornotemark[1]
\orcid{0000-0001-7290-068X}
\email{mgrunde@cs.washington.edu}
\affiliation{%
  \institution{University of Washington}
  \city{Seattle}
  % \state{Washington}
  \country{USA}
}

\author{Andrew McNutt}
% \authornotemark[1]
\email{amcnutt@cs.washington.edu}
\orcid{0000-0001-8255-4258}
\affiliation{%
  \institution{University of Washington}
  \city{Seattle}
  % \state{Washington}
  \country{USA}
}

\author{Jeffrey Heer}
\email{jheer@uw.edu}
\orcid{0000-0002-6175-1655}
\affiliation{%
  \institution{University of Washington}
  \city{Seattle}
  % \state{Washington}
  \country{USA}
}

\author{Tim Althoff}
\email{althoff@cs.washington.edu}
\orcid{0000-0003-4793-2289}
\affiliation{%
  \institution{University of Washington}
  \city{Seattle}
  % \state{Washington}
  \country{USA}
}

%%
%% By default, the full list of authors will be used in the page
%% headers. Often, this list is too long, and will overlap
%% other information printed in the page headers. This command allows
%% the author to define a more concise list
%% of authors' names for this purpose.
% \renewcommand{\shortauthors}{Trovato and Tobin, et al.}
\renewcommand{\shortauthors}{Gu et al.}

%%
%% The abstract is a short summary of the work to be presented in the
%% article.

%%%% 1-abstract.tex starts here %%%%

\begin{abstract}
Data analysis is challenging as analysts must navigate nuanced decisions that may yield divergent conclusions. AI assistants have the potential to support analysts in \textit{planning} their analyses, enabling more robust decision making. Though AI-based assistants that target code \textit{execution} (e.g., Github Copilot) have received significant attention, limited research addresses assistance for both analysis execution \textit{and} planning. In this work, we characterize helpful planning suggestions and their impacts on analysts’ workflows. We first review the analysis planning literature and crowd-sourced analysis studies to categorize suggestion content. We then conduct a Wizard-of-Oz study (n=13) to observe analysts’ preferences and reactions to planning assistance in a realistic scenario. Our findings highlight subtleties in contextual factors that impact suggestion helpfulness, emphasizing design implications for supporting different abstractions of assistance, forms of initiative, increased engagement, and alignment of goals between analysts and assistants.

\end{abstract}


%%
%% Keywords. The author(s) should pick words that accurately describe
%% the work being presented. Separate the keywords with commas.

%% A "teaser" image appears between the author and affiliation
%% information and the body of the document, and typically spans the
%% page.
\begin{teaserfigure}
  \includegraphics[width=\textwidth]{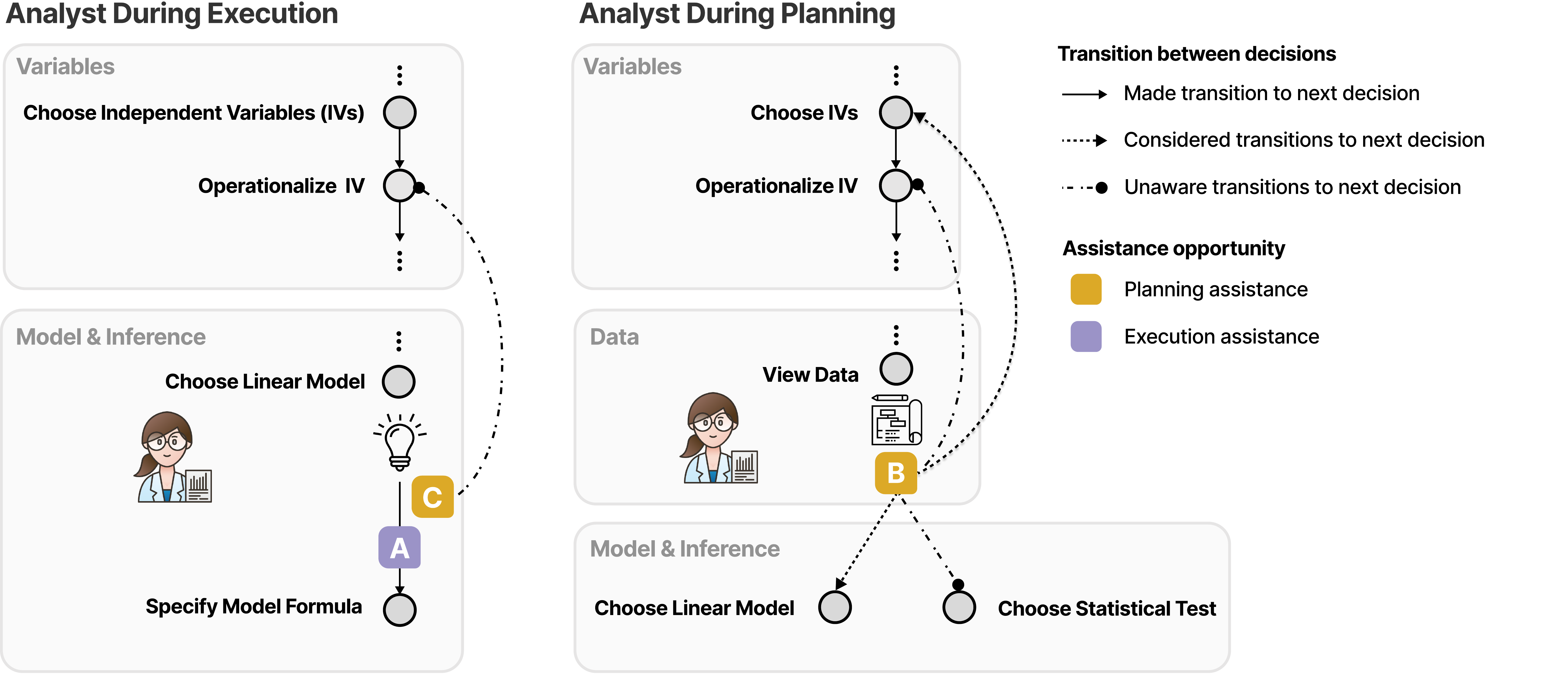}
  \caption{
  \textbf{Execution \textit{and} planning assistance during data analysis}. 
  During an ongoing analysis, analysts may be focusing on the \textit{execution} of an analysis decision (left) or \textit{planning} their next decisions (right). During execution, analysts have a clear intent of what they want to do (e.g., specify a model formula). Meanwhile, during planning, analysts are reasoning about potential decisions they are considering (\arrowConsidered ). 
  Existing systems powered by Large Language Models (e.g., ChatGPT and Github Copilot), focus on providing \textit{execution assistance}, helping the analyst carry out a decision (A). 
  % These systems either take the immediate programming context to recommend code or answer direct queries from the analyst.
  However, analysts can also benefit from \textit{planning assistance}. 
  Planning assistance helps analysts reason about the analysis decisions. 
  This can occur as analysts are explicitly planning their decisions (B) or even during execution when analysts are unaware of plausible alternatives (C).
}
  \Description{The figure shows the analysis workflow for analysts during execution (left) and analysts during planning (right). Analysis decisions are shown in grey circles with arrows pointing to the next analysis decision. On the left is the analyst during execution in which at the current step there is a light bulb indicating the analyst has an idea of what to do. There is an arrow pointing back to a prior step indicating an unaware transition to a prior decision. On the right is the analyst during planning. At the current step there is a pencil and paper icon indicating the analyst is considering their next analysis decision.}
  \label{fig:teaser}
\end{teaserfigure}

%%
%% This command processes the author and affiliation and title
%% information and builds the first part of the formatted document.

\maketitle

\section{Introduction}
\label{sec:intro}
Data analysis requires analysts to make numerous decisions regarding varied tasks, such as data collection, data wrangling, statistical modeling, and inference~\cite{Liu2019PathsEP}. 
Each choice involves \textit{analysis planning}, wherein analysts reason through the effects of potential decisions by synthesizing their results, prior experiences, and domain and statistical knowledge. 
Decisions made, their focus shifts to enacting these choices using the correct code with appropriate software tools, which we refer to as \textit{analysis execution}. After examining the outcome of their execution, the process repeats.
% Data analysis requires analysts to make many decisions across all phases of an end-to-end analysis (e.g., data collection, data wrangling, statistical modeling, inference, etc.)~\cite{Liu2019PathsEP}, and doing so involves analysis planning. \textit{Analysis planning} occurs when analysts reason through potential decisions to decide on subsequent analysis step(s). 
% or \tim{seems too narrow? what about when they forget a step and our assistant wants to bring that up?} 
% To inform their next decision, they synthesize current and past results with their own domain and statistical knowledge. Once analysts have made these decisions, their focus shifts to decision execution; \textit{analysis execution} involves implementing the decision using the correct code with appropriate software tools. 

% While execution and planning are both important, inadequate planning has significant negative ramifications.
% Researchers have identified poorly conducted analyses as a major contributor to the reproducibility crises \cite{Baker20161500SL}. Adding to the problem, analysis planning is challenging even to trained analysts. 
Inadequate analysis planning can complicate this flow and can cause analysts to risk making arbitrary and ill-founded decisions~\cite{Simonsohn2015SpecificationCD, Steegen2016IncreasingTT} for a host of reasons.
% foment \am{wc}unsubtantiate conclusions.
 % Inadequate analysis planning and execution can \am{wc}incur significant costs. 
% Lacking planning and sufficient execution skills, analysts risk making arbitrary and ill-founded decisions~\cite{Simonsohn2015SpecificationCD, Steegen2016IncreasingTT}.
% Analysts tendency ~\cite{Liu2019PathsEP, Liu2020UnderstandingTR} struggle to identify and address relevant decision points---potentially leading to unfounded conclusions. 
 Analysts often struggle to identify and address relevant decision points~\cite{Liu2019PathsEP, Liu2020UnderstandingTR}. 
 They may overfocus on low-level details, such as tuning hyperparameters, or miss high-level considerations like alternative statistical or mental models~\cite{Liu2020UnderstandingTR}.
 % , deprioritizing planning altogether. 
 They often center execution decisions tied to familiar workflows when approaches using other tools may be more applicable~\cite{Liu2020UnderstandingTR, Jun2021HypothesisFE}.
 % In this case, barriers to the execution of other decisions lead analysts to focus only on the most comfortable approaches. 
% 
% Since different decisions can lead to different conclusions, 
Further, the unconsidered flexibility in decision-making (and the biases inherent therein) contributes to the scientific reproducibility crisis~\cite{Baker20161500SL, Cockburn2020ThreatsOA, Aarts2015EstimatingTR}. 
% \jeff{Is this really a "trend"? Seems more to me to likely be a longstanding challenge we are now better putting into relief} \tim{+1. others call it a reproducibility crisis} 
For instance, given the same analysis task and dataset, analysts often derive divergent conclusions \cite{Silberzahn2018ManyAO, Breznau2022ObservingMR, Schweinsberg2021SameDD, Dutilh2018TheQO, BotvinikNezer2019VariabilityIT, Bastiaansen2019TimeTG, Menkveld2021NonStandardE}. 
% Furthermore, their inattention to key missed steps or alternative methods may make them overconfident in their findings. 
% All of this can lead to an insufficient \tim{misleading?} analysis, something researchers have identified as a major contributor to the reproducibility crisis in science \cite{Baker20161500SL}. \tim{can move the crisis up top of paragraph}

% and which leads to suboptimal decision making in industry and government. \ken{is it just industry? what about government?}.

% note to self: reproducibility crisis and divergent conclusions are two different but similar things
\textit{Data analysis assistants} seek to bridge these gaps by helping analysts execute \emph{and} plan their analyses. 
% \textit{Data analysis assistants} may improve analysts’ ability to design and enact decisions by offering execution \textit{and} planning assistance.
AI-based execution assistants (such as Copilot~\cite{githubcopilot}) help analysts implement decisions (Fig. \ref{fig:teaser}A) by suggesting code or tools. 
In contrast, planning assistance may help them reason about their decisions (Fig. \ref{fig:teaser}B and C), 
articulate hypotheses and mental models, and identify overlooked alternative decisions or rationales (Fig. \ref{fig:teaser} \arrowUnaware).
% Prior work~\cite{Liu2020BobaAA, Simonsohn2015SpecificationCD, Steegen2016IncreasingTT} indicates that systematically considering these alternative approaches could deepen analysts' understanding of the variations in outcomes inherent in their decisions. Hence, planning assistance can be seen as a pivotal step in promoting more theoretically principled, comprehensive, reliable, and therefore more robust analyses.
Systematically considering alternatives can deepen analysts' understanding of the outcome variations latent to decision-making in data analysis~\cite{Liu2020BobaAA, Simonsohn2015SpecificationCD, Steegen2016IncreasingTT}. 
% Systematically considering alternative approaches can deepen analysts' understanding of the variations in outcomes latent to their decisions~\cite{Liu2020BobaAA, Simonsohn2015SpecificationCD, Steegen2016IncreasingTT}. 
Therefore, planning assistance can be a transformative step towards more principled, reliable, and robust analyses.
% With quality assistance of both types, analysts can confidently select and accurately carry out their decisions. Altogether, such assistance may enable analysts to draw more reliable inferences. 
% 

 % LLMs are linguistic feed-forward machines that predict future text from past text \cite{Vaswani2017AttentionIA}. Thus, they are well suited to tasks in which the desired outcome is clear based on the textual context and are well-equipped for execution assistance. LLM-based code assistants like Github Copilot \cite{githubcopilot}recommend relevant code based on what the programmer has written so far (Fig. \ref{fig:quadrant}A and B). 
 % Given this functionality, LLM-powered code assistants have become increasingly popular in modern code environments. 
 % Similarly, LLM-powered analysis assistants can be tightly integrated into analysts' computational environments such as computational notebooks. 
% Similarly, LLM-based analysis assistants could in theory provide insightful recommendations for analysis planning.
 
\begin{figure}[t]
  \centering
  \includegraphics[width=0.92\linewidth]{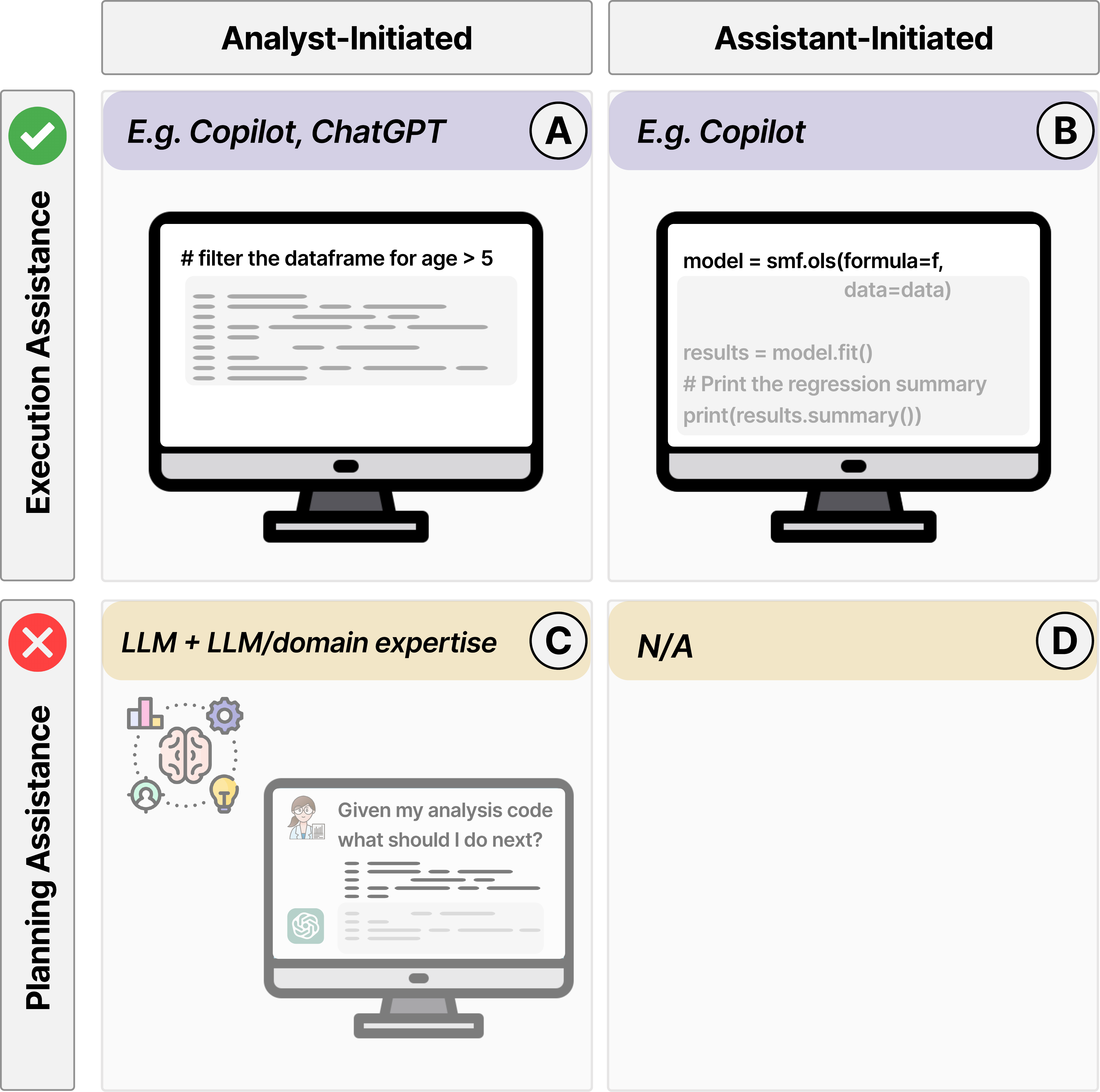}
  \caption{\textbf{Planning assistance has limited support and is unexplored}. Existing assistants, such as Github Copilot,  are well-suited for execution assistance, such as suggesting data wrangling or modeling code. 
  % Suggestions can be initiated by the analyst, such as through an explicit query (A), or by the assistant, based on the current context (B). While tools like ChatGPT may provide basic planning assistance (C), analysts can struggle to write effective prompts. Thus, there may be easier ways to initiate planning assistance or opportunities to rely on assistant initiation (D). However, it is unclear how to provide assistant-initiated planning assistance. Naive applications of LLMs alone cannot support this since they do not know the timing and relevant context for an assistant. Thus, they require selective context, higher-level scaffolding, targeted prompts, and triggers. 
  % In practice, planning assistance may be offered in conjunction with execution assistance to be easily adapted by the analyst.
  }
  \Description{The figure shows four quadrants. From left to right and top to bottom, the quadrants are labeled A, B, C, and D. The x-axis is labeled analyst-initiated and assistant-initiated. The y-axis is labeled execution assistance and planning assistance. The top row indicating execution assistance shows existing systems such as Github Copilot and ChatGPT. The bottom left quadrant shows analyst-initiated planning assistance which needs LLMs and LLM/domain expertise. The bottom right quadrant for assistant-initiated planning assistance show N/A for no current systems.}
  \label{fig:quadrant}
\end{figure}

The improved capabilities of large language models (LLMs) for coding, language, and visualization generation \cite{Brown2020LanguageMA, Bubeck2023SparksOA, Chen2021EvaluatingLL, Chowdhery2022PaLMSL, Dibia2023LIDAAT} offer an opportunity to make such planning \emph{and} execution data analysis assistants a reality.
Prior LLM-based assistants have explored practical approaches to execution assistance~\cite{Jiang2021GenLineAG, Vaithilingam2022ExpectationVE, Barke2022GroundedCH, McNutt2023OnTD, Ziegler2022ProductivityAO, Gu2023HowDA}, however planning remains unexplored (Fig.~\ref{fig:quadrant} bottom row).
% \am{finish sentence}.
% Prior research has explored the design and impacts of execution assistance from LLM-based code assistants \cite{Jiang2021GenLineAG, Vaithilingam2022ExpectationVE, Barke2022GroundedCH, McNutt2023OnTD, Ziegler2022ProductivityAO}. 
% However, additional challenges arise when considering the design of an assistant that offers both execution \textit{and} planning assistance.
% Existing assistants do not easily facilitate planning assistance (Fig.~\ref{fig:quadrant} bottom row): 

Effective planning suggestions can not be achieved by simply feeding the current analysis context into an LLM as it requires selective context, higher-level scaffolding, targeted prompts, and sensitive triggers.
The iterative nature of data analysis demands the nuanced interpretation of prior results, statistics, and domain knowledge to inform the subsequent decisions \cite{Grolemund2014ACI, Box1976ScienceAS, Jun2021HypothesisFE, Tukey1966DataAA}, which is typically unaccounted for in execution-only assistants (Fig. \ref{fig:quadrant}A and B). 
Synthesis of such decisions may require backtracking to a prior choice and deviating from the current approach (Fig.~ \ref{fig:teaser}).
Authoring effective prompts can be difficult~\cite{ZamfirescuPereira2023WhyJC, Mishra2022HELPMT}, and the right context to provide to not cause hallucinatory analyses~\cite{mcnutt2020surfacing} may be hard to identify.
Finally, data analysis is a non-linear activity that involves iterated and inter-woven actions of typing, reading, and reflecting---making the optimal timing of suggestions less clear than in software engineering.

Yet, essential questions about the design of planning assistants remain unanswered.
In this work, we explore the design of such AI-based  assistants.
% , in addition to more familiar execution help. 
% Since execution assistance is relatively well-explored and planning assistance less well so, we focus on understanding the interfaces, interactions, and processes for effectively incorporating planning assistance into analysts' workflows. 
To this end, we consider the following research questions:
\begin{description}
\item[RQ 1 - Scope:] What types of specific planning assistance content could an assistant provide?
% As there are many situations that may benefit from analysis planning, what content of types of specific planning feedback should an assistant provide?
\item[RQ 2 - Helpfulness:]
% As there are different methods of interaction (e.g., user-initiated vs. assistant-initiated), forms of assistance (i.e., planning vs. execution assistance), and specific suggestions under each type of assistance (e.g., modeling or data wrangling during planning), 
% What types of suggestion content do analysts find helpful? Are there other characteristics that make a helpful suggestion?
What data analysis suggestions do analysts find helpful and under what circumstances?
\item[RQ 3 - Impact:] How do suggestions impact analysts’ workflows? To what extent do analysts prioritize their own ideas or adopt suggestions and explore alternative approaches?
% Do analysts take the suggestions over their own ideas and consider alternative approaches more frequently? \tim{unsure about second sentence - what else do we consider? perhaps whenever suggestions DO NOT get considered? right now this example question seems to "bias" our inquiry. Can you make it broader or more balanced to what you do the most or what readers to  take away?}
\end{description}

To characterize assistant suggestion content (\textbf{RQ 1}), we conducted a \litreview review and critically assessed data from a crowd-sourced analysis study~\cite{Silberzahn2018ManyAO}, wherein 29 analyst teams separately analyzed the same dataset to answer the same research question (Sec. \ref{sec:suggestions}). 
We used these findings to develop \textit{categories of assistance content} across points in the end-to-end analysis process (Table \ref{tab:suggestions}).
% By studying the literature, we identify challenges analysts face and broad categories of assistance opportunities. Likewise, by observing variations in independent analysis approaches in the crowd-sourced analysis study, we identify situations where analysts would benefit from being offered alternative approaches or reminders for key analysis steps. This process helped us develop \textit{categories of assistance content}
% % \tim{this sounds broader than the "suggestion content" mentioned in the abstract} 
% across points in the end-to-end analysis process (Table \ref{tab:suggestions}).

To observe analysts' preferences for (\textbf{RQ2}) and effects of (\textbf{RQ3}) different suggestions, we conducted a \woz study, with 13 practicing data analysts (Sec. \ref{sec:labstudy})---our first contribution\footnote{The study materials are available in the supplementary material and online on OSF at \osfURL.}. We designed the study around the scenario considered in the crowd-sourced study~\cite{Silberzahn2018ManyAO} that we looked at in our previous step.
Our participants worked through the same dataset and task while receiving suggestions from a behind-the-scenes experimenter acting as a wizard.
% , which allowed us to simulate high-quality and relatively low-latency feedback. 
% to elicit discussion about the helpfulness of a suggestion and observe factors influencing the analyst-assistant interaction. 
% Instead of evaluating the quality of an analysis assistant that is likely to be imperfect, we focus on studying how these assistants should be designed in the first place. We build on prior works on the design of code-assistants in computational notebooks \cite{McNutt2023OnTD}, and guidelines in Human-AI interaction \cite{Amershi2019GuidelinesFH, Horvitz1999PrinciplesOM}.
To facilitate our study, we developed a JupyterLab extension that allowed the wizard to surface suggestions to the participant as they used the notebook in a familiar manner\footnote{To support future notebook assistant interfaces, we release the code for this interface at \githubURL.}. 
% to act as planning assistant (aided by contemporary LLMs). 
% For the study, we developed a prototype of an analysis assistant’s interface and associated interactions. 
% We also created a wizard protocol that was not limited by current assistants and LLMs to elicit findings and design implications for LLM-based analysis assistants of the near future; we used current LLMs to generate and iteratively refine suggestions. 
This allowed the wizard (aided by LLMs) to synthesize suggestions and manipulate the selection, quality, and timing of recommendations (Fig.~\ref{fig:woz}). 
We synthesize the findings of our study (Sec. \ref{sec:results}) into a set of design guidelines (Sec.~\ref{sec:discussion}) to align analysis intent and goals, support varying levels of suggestion assistance, and increase analyst understanding and engagement with the assistant's suggestions---our second contribution.
% Among study findings (Sec. \ref{sec:results}), 
We observed that the data analysis process can benefit from diverse forms of planning assistance, but only with careful consideration of assistance context. 
Additional factors---such as the timing of suggestions or variations in analysts' statistical and domain backgrounds---played a critical role in a suggestion's perceived helpfulness. Finally, we found differences in goals between the analyst (i.e., speed and completion) and planning assistance (i.e., methodical and robust planning).
% By characterizing these usability concerns, we hope to make planning-equipped data analysis assistants practical and effective, and thereby make data analysis more robust and reliable.

By characterizing the usability of planning-equipped analysis assistants, we seek to make data analysis more robust and reliable by enabling richer and more useful assistants.
% \am{by doing this we change the world or whatever}

% \tim{missing, say that you conduct the woz study, and what comes after (say thematic analysis of qualitative feedback etc etc.}

% The primary contributions of this paper include: 
% % The paper contributes the following:
% \begin{enumerate}

% % \item A categorization of data analysis planning suggestion content 
% % \item An open-source data analysis assistant interface prototype integrated into JupyterLab
% \item Findings from a \woz study that reveal multiple interacting factors that influence the helpfulness of data analysis planning assistance  
% \item Design guidelines for incorporating planning assistance
% % and execution assistance \tim{bit confusing to now see execution here again while we highlighted planning before. IIRC we do not highlight that these are intertwined in intro} 
% in a data analysis assistant (Table~\ref{tab:design_implications})

% \end{enumerate}

% In addition, we release the code for our assistant interface prototype used in the \woz study.\footnote{Github}

%%%% 2-intro.tex ends here %%%%

%%%% 3-related-work.tex starts here %%%%

\section{Background and Related Work}
\subsection{Planning in Data Analysis}
\label{sec:bg_planning}
% In an interview study with professional data analysts, Alspaugh et al. \cite{Alspaugh2019FutzingAM} describe how analysts often engage in activities related to planning their analysis approach. 
Analysts often engage in planning activities~\cite{Alspaugh2019FutzingAM}. Such activities are closely related to cognitive theories of \textit{sensemaking}, whereby analysts seek and integrate new observations to build mental models~\cite{Pirolli2007TheSP, Russell1993TheCS, Klein2007ADT} that inform their decisions for collecting and analyzing data to elicit findings; these findings, in turn, inform their mental models in an iterative back-and-forth.
Thus, data analysis planning can be highly iterative and often requires revisiting decisions at various points in the analysis~\cite{Jun2021HypothesisFE, Box1976ScienceAS, Tukey1966DataAA}. 
Moreover, these processes are integral to reduce human biases in decision-making and help form robust conclusions \cite{Grolemund2014ACI}. In this work, we study how sensemaking activities in analysis planning can be improved with an AI assistant.

Similarly, analysis planning relates closely to \textit{multiverse analysis}~\cite{Steegen2016IncreasingTT, Simonsohn2015SpecificationCD}, a statistical analysis paradigm whereby analysts consider, specify, and report all reasonable decisions and combinations of decisions. Through assessing potentially thousands of analyses based on these choices, a multiverse approach highlights how different decisions can influence final outcomes. Despite the importance of analysis planning, empirical studies observed analysts struggle to reason about their analysis decisions \cite{Liu2019PathsEP, Liu2020UnderstandingTR}. Often, they are unaware of potential alternative decisions in the end-to-end data analysis process.
% (i.e., the phases of data collection, data wrangling, statistical modeling, and evaluation). 
Our work aims to guide analysts in recognizing, understanding, and reflecting on these overlooked but important decisions. 

Prior work developed systems to simplify the planning of statistical models and statistical tests~\cite{Jun2022TisaneAS, Jun2019TeaAH}. Data-driven systems (like Lodestar~\cite{Raghunandan2022LodestarSI}, EDAssistant~\cite{Li2021EDAssistantSE}, and ATENA \cite{El2020AutomaticallyGD}) add structure to analyses and offer recommendations for data science and exploratory data analysis (EDA) workflows. However, these systems focused on specific parts of the analysis process (e.g., modeling or EDA) or based their recommendations on predicting the next steps in an analysis. In contrast, \textit{we emphasize analysis planning for the entire end-to-end analysis and as a highly iterative process}. 

Specifically, we aim to support analysts in considering alternative decisions and revisiting prior decisions as new insights emerge. We view LLM-supported data analysis assistants as a promising solution. Furthermore, these assistants could be integrated with existing multiverse analysis systems~\cite{Liu2020BobaAA, Sarma2021multiverseMA, Gu2022UnderstandingAS}. % to help analysts conduct multiverse analyses. 
Thus, as planning assistants help analysts consider and structure alternative decisions, multiverse tools help them author, execute, and evaluate these decisions within a ``multiverse'' of alternative analyses.

For these benefits to be realized, we must learn what specific feedback analysts need and under what circumstances assistance would be most useful to them. To begin exploring factors that influence favorable suggestions, we first categorize suggestion content (Sec. \ref{sec:suggestions}). Using this categorization, we develop a prototype interface and conduct a \woz study (Sec. \ref{sec:labstudy}) to investigate the helpfulness and impact of such suggestions.

% With LLMs' capability to generate coherent code and text, the assistant can offer information on relevant domain knowledge and statistical concepts, along with precise code implementation.

% \tim{again, this is a nice similarity but it is critical for you do highlight significant differences, additional contributions and why they are super important in all of the sections if not paragraphs here.}

\subsection{LLM-based Assistants and Limitations for Planning Assistance}
\label{sec:llm_background}

% Modern programming assistants like Github Copilot utilize the strong text generation abilities of Large Language Models (LLMs). These LLMs are trained on extensive text corpora, often sourced from the web \cite{Brown2020LanguageMA, Zhang2022OPTOP}. 

LLMs trained on large-scale code corpora \cite{Chen2021EvaluatingLL, Chowdhery2022PaLMSL, Fried2022InCoderAG, Nijkamp2022ACP, Li2022CompetitionlevelCG} have shown proficiency in learning programming concepts, solving programming challenges \cite{Chen2021EvaluatingLL}, and covering various domains like web development \cite{Zhang2023PracticesAC} and data science problems \cite{Lai2022DS1000AN, Chandel2022TrainingAE}. While these LLMs excel at precise code implementation tasks (e.g., \textit{Consider a (6,7,8) shape array, what is the index (x,y,z) of the 100th element?} \cite{Lai2022DS1000AN}), this work focuses on situations where the problem statement and methods to apply are less clear, unspecified by the user, or at a higher level of abstraction. 
% \tim{is that it? or could it just be broader, or specified on a much higher level of abstraction?}

% Modern programming assistants such as Github Copilot are enabled by the strong text generation abilities of Large Language Models (LLMs). These LLMs are trained on vast corpora of text, often taken from the web \cite{Brown2020LanguageMA, Zhang2022OPTOP}. In particular, LLMs trained on large-scale code corpora \cite{Chen2021EvaluatingLL, Chowdhery2022PaLMSL, Fried2022InCoderAG, Nijkamp2022ACP, Li2022CompetitionlevelCG} have been shown to learn programming concepts, solve programming challenges \cite{Chen2021EvaluatingLL}, and cover a range of task domains from web development \cite{Zhang2023PracticesAC} to data science problems \cite{Lai2022DS1000AN, Chandel2022TrainingAE}. However, the tasks these LLMs address all involve very specific problem statements asking for precise code implementation (e.g., \textit{Consider a (6,7,8) shape array, what is the index (x,y,z) of the 100th element?} \cite{Lai2022DS1000AN}). In contrast, this present work addresses common situations when the problem statement is less clear or even unspecified by the user. % \tim{if they can do DS problems -- why are they still insufficient or not well enough understood as we claim here?}

LLMs are trained to predict the next token in a code/text sequence and thus form the back-end of many popular code assistants that offer execution assistance \cite{Murali2023CodeComposeAL, githubcopilot, googleblog2022, codewhisperer}. Note that we distinguish between an LLM and an assistant; the latter is a \textit{complete system} that acts as the interface between the user and the LLM. Existing code assistants typically pass the programmer's context (e.g., code, comments, etc.) to the LLM to then recommend entire statements or blocks of code. In this interaction, programmers can either initiate execution assistance with explicit comments and triggers (Fig. \ref{fig:quadrant}A) or passively let the assistant auto-complete their code (Fig. \ref{fig:quadrant}B). However, as discussed earlier  (Sec.~\ref{sec:intro}), assistants offering data planning assistance cannot naively pass the current context to an LLM given the iterative and non-linear nature of analysis planning. 
% \tim{say why -- the why is more important than the  fact you explained it before}\ken{we went thorough in this in intro and find that we would be repeating ourselves here. This was part of the densification discussed earlier}. 
This work explores the design of an LLM-supported complete system for planning assistance. 

% At their core, LLMs are trained to predict the next token in a code/text sequence. Based on this idea, LLMs form the back-end of many popular code assistants that offer execution assistance \cite{Murali2023CodeComposeAL, githubcopilot, googleblog2022, codewhisperer}. Note, we make the distinction between an LLM and an assistant, which is a full-fledged system around the LLM acting as the interface between the user and the LLM. With existing code assistants, they typically pass the programmer's context (e.g., code, comments, etc.) to the LLM to then recommend entire statements or blocks of code. Within this interaction, programmers can either initiate execution assistance with explicit comments (Fig. \ref{fig:quadrant}A) or passively let the assistant auto-complete their code (Fig. \ref{fig:quadrant}B). In data analysis, however, suggestions to help reason about analysis decisions (i.e., planning assistance) may require backtracking to points much earlier in the analysis and can deviate from the analyst's immediate context \cite{Liu2019PathsEP, Grolemund2014ACI}. As a result, unlike existing code assistants, data analysts assistants when offering planning assistance cannot naively pass the current context to an LLM. \tim{add a last sentence about what your work adds here} \tim{move figure closer to here}

% The assistants include additional interfaces and interactions between the user and the LLM.

More recently, general-purpose LLMs, such as InstructGPT \cite{Ouyang2022TrainingLM} and GPT-4 \cite{OpenAI2023GPT4TR}, 
have reached a level of performance that is now deployable in various contexts and domains~\cite{Bubeck2023SparksOA}.
% have achieved human-level \tim{i hesitate to call it that since it's easily misleading. is there a good other way to specify the level of performance?} performance across many difficult tasks in various domains \cite{Bubeck2023SparksOA}.  
% They are fluent in human-style conversations and can naturally engage in multi-turn conversations with the user.
The assistant ChatGPT~\cite{ChatGPT}, supported by LLMs, has exploded in popularity \cite{statista}. In the context of data analysis, ChatGPT can offer basic planning assistance, integrating broader domain expertise, software libraries, and statistical knowledge with code when given precise instructions (e.g., \textit{Show me the correlation between Height and Weight in a scatter plot})~\cite{Cheng2023IsGA}.

However, ChatGPT requires users to actively prompt it for information (Fig.~\ref{fig:quadrant}C). Prior work found that non-AI experts often struggle to write effective prompts \cite{ZamfirescuPereira2023WhyJC} and experience significant cognitive load~\cite{Jiang2022PromptMakerPP}. Data analysis planning can further exacerbate this issue since analysts must synthesize information from various points in the analysis to pass to the assistant. Thus, the burden is on the analyst to decide what they want, where to find relevant context in a likely messy analysis \cite{Head2019ManagingMI, Kery2018TheSI, Rule2018ExplorationAE}, and craft a prompt or series of prompts. In addition, since analysts are often unaware of the entire space of analytical decisions, they may also need planning assistance they do not directly ask for. However, current assistants like ChatGPT do not naturally support assistant-initiated planning assistance (Fig. \ref{fig:quadrant}D). 
Further, since analysts may be alternating between planning and execution, the optimal timing of planning assistance is unclear. 
% Given all these challenges, current LLMs cannot be naively employed for planning assistance. 

Our work aims to understand the design of data analysis assistants that involve all forms of assistance and initiation. Since planning assistance presents new and largely unexplored challenges, we focus our efforts here. Notably, we monitor analysts' reactions to an assistant that offers new planning information that possibly differs from their current analysis context (i.e., analysis code and notes). In accomplishing this, we also explore an assistant that raises suggestions without direct invocation from the analyst.

\subsection{Designing for Human-AI Interaction}
In general, designing AI-based user-facing systems is challenging since designs must face issues that include explainability \cite{Liao2020QuestioningTA, Kim2022HelpMH}, trust \cite{Kunkel2019LetME, Liao2022DesigningFR, Toreini2019TheRB}, user control \cite{Shneiderman2020HumanCenteredAI, Horvitz1999PrinciplesOM}, and user expectations \cite{Luger2016LikeHA, Kocielnik2019WillYA}. As a result, there is a rich history of literature on design guidelines for Human-AI interaction to address these challenges \cite{Horvitz1999PrinciplesOM, Amershi2019GuidelinesFH}.
Similar practitioner-facing guidelines have also been released in large companies \cite{PAIRGuidebook, appleML, IBMDesignAI}. 

% Horvitz's foundational work on mixed-initiative user interfaces  outlined principles for having humans and automated services interact effectively \cite{Horvitz1999PrinciplesOM}. Twenty years later, Amershi et al. \cite{Amershi2019GuidelinesFH} built on Horvitz's set of principles and formulated 18 generally applicable guidelines for Human-AI interaction. 
% In creating these guidelines, Amershi et al. explored AI products and guidelines in industry and academic literature about AI design. They also went through multiple phases of evaluation and refinement to create the final set. \tim{maybe unnecessary detail} 
% Similar practitioner-facing guidelines have also been released in large companies \cite{PAIRGuidebook, appleML, IBMDesignAI}. 

Though these resources address common challenges in Human-AI interaction design, designing a system still requires support for domain-specific examples and proper problem formulation \cite{Yildirim2023InvestigatingHP}.
In other words, these resources do not address the specific Human-AI interaction challenges surrounding the design of data analysis assistants (e.g., the timing of assistant-initiated planning assistance).
% \tim{can we say that they don't and perhaps provide evidence or an example?}
Our work focuses exclusively on identifying precise goals and problems for designing analysis assistants. Based on findings from a \woz study, we propose design guidelines specifically for developing these assistants (Table~\ref{tab:design_implications}). 

% With respect to designing assistants for data analysis, McNutt et al. \cite{McNutt2023OnTD} studied the design of AI-powered code assistants in computational notebooks. Their work focused on the computational notebook as a medium to provide execution assistance. They conducted an interview study and presented analysts with slide-based prototypes of interfaces. 
With respect to designing analysis assistants, McNutt et al. \cite{McNutt2023OnTD} studied computational notebooks as a medium to provide AI-based execution assistance for data scientists. They conducted an interview study and presented participants with slide-based prototypes of interfaces.
In contrast, our work concentrates on the open challenges in understanding the scope of (Sec.~\ref{sec:suggestions}) and developing for (Sec.~\ref{sec:discussion}) planning assistance.
% , we take extra consideration into studying planning assistance. \tim{extra consideration sounds very vague...} 
Instead of focusing on the medium where the assistant resides, we focus on identifying factors that make assistance helpful.

%%%% 3-related-work.tex ends here %%%%

%%%% 4-suggestions.tex starts here %%%%

\section{Categorizing Suggestion Content}
\label{sec:suggestions}

\begin{table*}
  \small
  \begin{tabular}{p{2.6cm}p{13.5cm}p{0.0cm}}
    \toprule
    \textbf{Suggestion Category} & \textbf{Example} &   \\
    \midrule
     Domain Background & \textbf{Similar Analyses Done in the Past} \newline
    The question of whether skin tone influences the way referees make decisions in soccer games has been a topic of research. 
    "Racial Bias in the Allocation of Fouls in Soccer" by Price and Wolfers (2010) is a representative study...
    % One of the most well-known studies in this area is "Racial Bias in the Allocation of Fouls in Soccer" by Price and Wolfers (2010).  
    & \\
    \midrule
     Data Wrangling \newline Assistance & \textbf{Reminder - Check for Missing Data} \newline
    Always check for missing data values in your DataFrame before performing any data analysis or machine learning tasks. Missing data can cause unexpected errors and affect the accuracy of your results. 
        \begin{lstlisting}[caption={}, label={lst:pythoncode}, belowskip=-0.8 \baselineskip]
# Use boolean indexing to filter the DataFrame and show only the rows with missing values
rows_with_na = df[df.isna().any(axis=1)]
rows_with_na.head() #...\end{lstlisting} & \\
    \midrule
     Conceptual Model \newline Formulation & \textbf{Other Variables that Might Influence Red Cards: Age Variable Rationale}  \newline
    Beyond just skin tone, we can consider other variables that might affect the analysis. As \emph{birthday} is in the dataframe, we might want to control for age, which is a common practice in research studies, since age can affect various aspects of behavior and cognition. One reason to control for age is to account for the tendency of impulsivity, which may be associated with receiving red cards, which decreases with age... &   \\
    \midrule
     Operationalizing \newline Constructs & \textbf{Skin Tone Operationalization}  \newline
    Right now, skin tone rating is represented by \emph{rater1} and \emph{rater2}. The \emph{rater1} and \emph{rater2} variables could be averaged and rounded to the most central value. We could also create a binary variable based on a rating threshold. This way, we could work with only one variable for player skin tone. It is up to you to decide which decision to make to ensure the validity and reliability of your results.
    This code shows a template for representing skin tone...
    &  \\
    \midrule
     Choosing the \newline Statistical Model & 
    \textbf{Using a Poisson Model}  \newline
    Here, we want to investigate the relationship between a player's skin tone and the number of red cards they receive from referees. Since the outcome variable, i.e., the number of red cards received by a player is a count variable (i.e., it is a non-negative integer), it might make sense to consider a Poisson model to analyze this relationship...
    &   \\
    \midrule
     Model Results \newline Interpretation & \textbf{Results Interpretation}  \newline
    The regression suggests that skin tone and the number of goals are significant predictors red cards. The R-squared value of 0.005 indicates that the model explains only a small proportion of the variance in the \emph{redCards} variable... 
    & \\
    \midrule
     High-Level Planning & \textbf{Consider Including Covariates}  \newline
Covariates are additional variables that can affect the relationship between the predictor variable and the outcome variable in an analysis. To accurately understand the relationship between the predictor and outcome variable, it is important to consider the effects of these covariates. Including covariates in an analysis can help to control for potential confounding variables... &  \\
    \midrule
     Execution Assistance & \textbf{Answering: Is the number of red cards associated with some unique referee ID?}  \newline
    Here is some code to answer this question.
    % \texttt{\# Group the data by unique referee IDs}
            \begin{lstlisting}[caption={}, label={lst:pythoncode}, belowskip=-0.8 \baselineskip]
referee_groups = df.groupby('refNum') #...\end{lstlisting} & \\
    \bottomrule
    
  \end{tabular}
  \caption{Categories of suggestions informed by a literature review and a behavior-driven analysis of a crowd-sourced data analysis study. With the exception of \textit{high-level planning} and \textit{execution assistance}, the dimensions correspond to assistance during the stages in an end-to-end analysis. \textit{High-level planning} involves more general considerations.}
  \label{tab:suggestions}
\end{table*}

To identify the circumstances in which planning suggestions are helpful (\textbf{RQ 2}), we first establish categories of relevant suggestion content (\textbf{RQ 1}). In our subsequent \woz study (Sec.~\ref{sec:labstudy}), we use this categorization to derive suggestions and observe analysts' reactions and preferences.
% under a realistic analysis setting. 
% 
We conducted a literature review 
% \tim{conduct a literature review? given that we end up w six papers and do not conduct a formal, comprehensive lit review I think we should try to find a different name for this. How about "theory driven"?} 
to identify existing challenges data analysts face. We then examined results from a crowd-sourced analysis study, where multiple independent analysts made decisions given the same data and research question. This helped us understand variations in  decision-making, determine points where alternative approaches would be plausible, and observe what these approaches might look like. 
% From studying a crowd-sourced analysis study, we build specific, highly-relevant examples of planning suggestions. 
% Our investigation offers a categorization of data analysis assistance content. 

 % \tim{readers may not know what this means. you need to explain that this is an ideal scenario where you observe the process and decision making of multiple independent analysts and that this helps to understand the variation in decision making and what alternative approaches are plausible. }
\subsection{Literature Review}
We began by identifying works that study the data analysis process and present opportunities for assistance. We searched Google Scholar with the keywords "sensemaking data analysis" from which we iteratively snowball sampled~\cite{Wohlin2014GuidelinesFS}. 
% \tim{would everyone understand that this is presumably done via references?}. 
We sought papers that discussed the end-to-end data analysis process and challenges suitable for a text-based assistant. 
% Because planning assistance is less understood and often has no invocation/guidance from the analyst, we direct our efforts toward this.
 Therefore, we did not focus on problems with specific systems \cite{Shrinivasan2008SupportingTA, Kang2012ExaminingTU} or address common analysis challenges related to software and debugging \cite{Kandel2012EnterpriseDA, Chattopadhyay2020WhatsWW, Gu2022UnderstandingAS}, communication \cite{Zheng2022TellingSF, Wang2021DocumentationMH}, and data provenance \cite{Kery2018TheSI, McNutt2023OnTD, Xu2015AnalyticPF}.

Our review resulted in 15 publications that addressed data analysis in various domains (e.g., food delivery, education, medical imaging, etc.), analysis tasks, and groups of data workers (e.g., enterprise workers, researchers, college students, etc.). From these works, we conducted thematic analysis on excerpts describing data analysis challenges and identified broad categories of assistance. 
 % \jeff{Only six papers? Can we more clearly indicate how we identified these and ruled out any others?} 
 These included: \textit{domain expertise information} \cite{Jun2021HypothesisFE, Liu2020UnderstandingTR, Wongsuphasawat2019GoalsPA, Boukhelifa2017HowDW, Wild1999StatisticalTI}, \textit{additional statistical knowledge} \cite{Jun2021HypothesisFE, Liu2019PathsEP, Liu2020UnderstandingTR}, \textit{guidance for statistical modeling} \cite{Jun2021HypothesisFE, Huber2011DataAW, Wild1999StatisticalTI, Bissell1988ProblemSA, Grolemund2014ACI}, \textit{guidance for analysts' hypothesis derivation} \cite{Jolaoso2015TowardAD, Battle2019CharacterizingEV, Kang2011CharacterizingTI, Alspaugh2019FutzingAM}, \textit{alternative decisions} \cite{Liu2019PathsEP, Liu2020UnderstandingTR, Kale2019DecisionMakingUU}, and data analysis specific \textit{execution assistance} \cite{Raghunandan2022CodeCE, Liu2020UnderstandingTR, Wongsuphasawat2019GoalsPA, Jun2021HypothesisFE, Alspaugh2019FutzingAM}.  
 We then refined these categories by comparing independent analyses from a crowd-sourced analysis study, described next.

\subsection{Behavior-Driven Analysis of a Crowd-Sourced Data Analysis Study}
\label{subsec:crowdsourced_study}
In crowd-sourced data analysis studies \cite{Silberzahn2018ManyAO, Breznau2022ObservingMR, Schweinsberg2021SameDD, Dutilh2018TheQO, BotvinikNezer2019VariabilityIT, Bastiaansen2019TimeTG, Menkveld2021NonStandardE}, numerous independent teams of analysts conduct analyses answering the same research question with the same dataset. Though originally intended by authors to study variations in analytical choices and how they impact analysis conclusions, these studies also offer high-fidelity data about code and rationales behind analysts' decisions. 

In this work, we chose a crowd-sourced study \cite{Silberzahn2018ManyAO} that addresses the research question: \textit{Are soccer players with a dark skin tone more likely than those with a light skin tone to receive red cards from referees?} The study provides a real-world dataset for answering the research question and includes analyses from 29 teams of analysts. The question and dataset are of moderate practical difficulty and complexity, encouraging different but reasonable analytical approaches. Furthermore, the domain of the study, soccer, is approachable to general audiences, reducing the likelihood of substantial differences in analysts' domain expertise.

We analyzed the soccer dataset for alternative approaches to an analysis decision at any point in the analysis process (e.g., data wrangling, statistical modeling, inference, etc.). To do so, we looked for decisions with high variance between analysis teams (e.g., choosing specific covariates).
% \mgm{What does this mean -- us analyzing for alternative approaches? Alternative to what?} 
From these decisions, we observed approaches (e.g., choosing player position as a covariate) that could help introduce other analysts to alternative approaches and mental models. These decisions informed categories of suggestions that aid in the various stages of the end-to-end analysis process.

These categories both corroborate and make more specific those outlined in the \litreview review. For instance, \textit{guidance for statistical modeling} from the \litreview review includes two aspects we uncovered in the crowd-sourced study: \textit{conceptual model formulation} and \textit{operationalizing constructs}. Likewise, \textit{alternative decisions} captures both \textit{high-level planning} (which helps analysts consider broader analysis decisions) and more specific planning decisions better summarized by the other categories. 
% We discuss different abstractions of planning assistance in Section~ \ref{sec:discussion_helpful}. 
From studying independent analyses of the crowd-sourced data, new categories, such as \textit{data wrangling} and \textit{model results interpretation}, also became apparent.

\begin{figure*}[t!]
  \centering
  \includegraphics[width=0.89\linewidth]{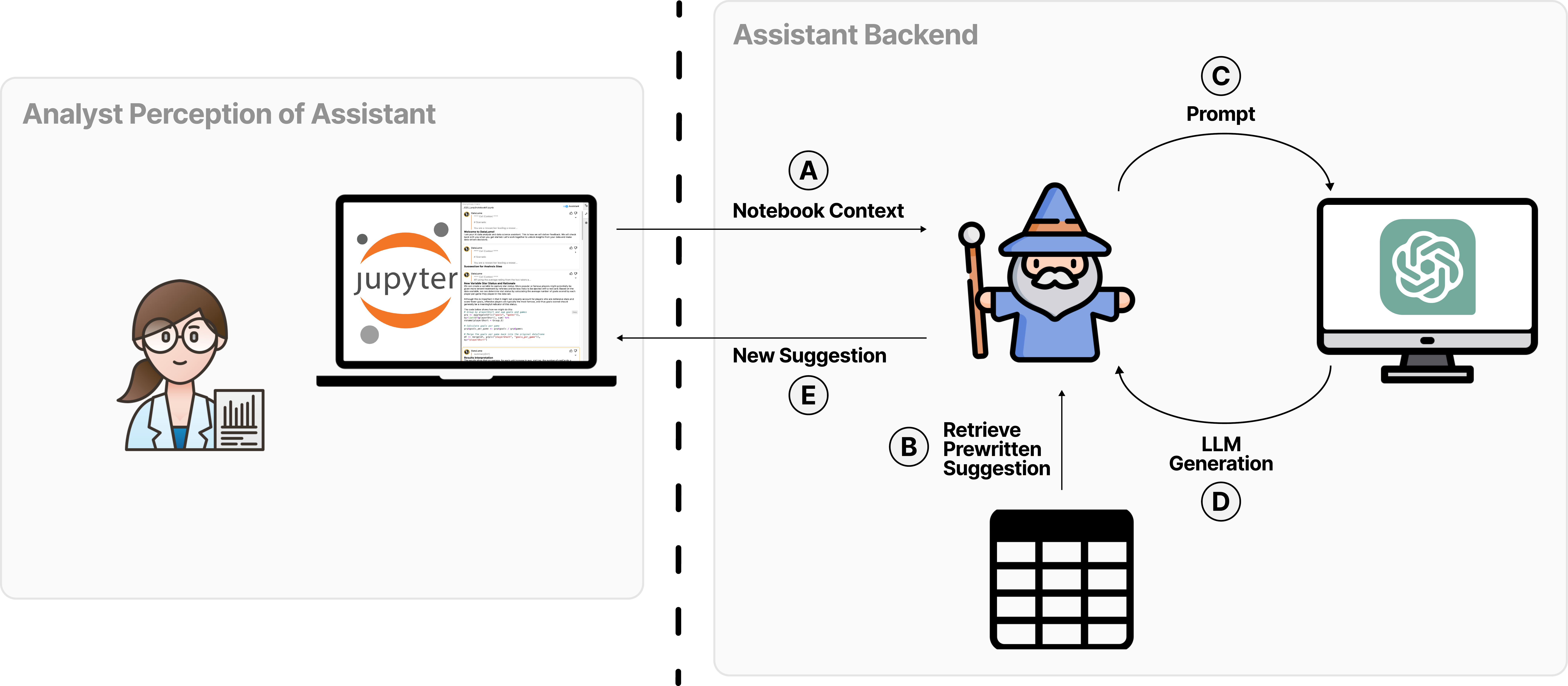}
  \caption{In our \woz study, analysts work with a JupyterLab notebook containing the assistant interface (left). They are unaware of the existence of a human wizard, located in a separate room, operating the assistant backend (right). The wizard uses a general purpose LLM and integrates information prepared before the study. The wizard observes the analyst's notebook for opportunities for assistance and obtains the relevant notebook context (A). Next, the wizard either retrieves a prewritten suggestion developed prior to the study (B) or prompts the LLM (C) to generate a relevant suggestion (D). The wizard may further refine the suggestion (e.g., shorten its length) with additional prompting of the LLM (C-D Loop). Finally, after creating the suggestion, the wizard presents the suggestion back to the analyst (E). In this process, the wizard acts as the interface between the analyst and the LLM, deciding on the content and timing of the suggestion. 
  % \tim{love the aesthetics! the dashed line has shading and nothing else does. I suggest to remove the shading for visual consistency}
  }
  \Description{The figure shows the analyst perception of the assistant on the left. This is an analyst icon with a JupyterLab notebook. The right side shows the assistant backend on the right consisting of a wizard icon, a monitor icon with ChatGPT and a spreadsheet icon. The figure also consists of the following arrows:
  1. The JupyterLab notebook to the wizard (labeled A)
  2. The spreadsheet to the wizard (labeled B)
  3. The wizard to the ChatGPT icon (labeled C)
  4. The ChatGPT icon to the wizard (labeled D)
  5. The wizard to the JupyterLab notebook (labeled E)
  }
  \label{fig:woz}
\end{figure*}
\subsection{Results}
 A summary of our suggestion categorizations is shown in Table \ref{tab:suggestions}, which includes examples we later developed for our \woz study (Sec.~\ref{sec:labstudy}). Full examples of suggestions in each category are also in the appendix. The \litreview review spanned domains and tasks, covering broad categories of assistance (e.g., \textit{guidance for statistical modeling}). 
% and guidelines (i.e., incorporate relevant domain and statistical knowledge) for crafting suggestions. 
Examination of the crowd-sourced study then helped facilitate more specific and new categories of help aimed at critical steps in the end-to-end analysis process (e.g., \textit{choosing the statistical model}). All categories except \textit{execution assistance} are \textit{intended} as planning assistance, helping the analyst reason about and consider alternative analysis decisions. Likewise, while the categories may not cover every possible helpful suggestion, based on our research, we expect high coverage. 

Overall, these categories encapsulate the collective approaches and rationale of expert-written analyses. Thus, they are likely to be helpful for analysts working on a variety of research problems. 
However, it is unclear to what degree suggestions from these categories would be helpful and what factors influence this judgment. To answer this, we conducted a \woz study.

%%%% 4-suggestions.tex ends here %%%%

% \input{main-design-motivation}

%%%% 5-woz-new.tex starts here %%%%

\section{\woz Study}
\label{sec:labstudy}
We conducted an exploratory lab study using a \woz protocol~\cite{Maulsby1993PrototypingAI, Dahlbck1993WizardOO} to understand the types of suggestions analysts prefer and how these suggestions might affect their workflow. 
In our study, participants interacted with a data analysis assistant housed in a customized JupyterLab interface which, unbeknownst to them, was controlled by a human ``wizard''. 
Our wizard (Fig. \ref{fig:woz}) simulated an LLM-based data analysis assistant. 
The wizard managed the LLM interaction manually, drawing from a predefined list of answers to make planning and execution suggestions while participants carried out a data analysis task.
% since analysts understandably may not have known how and when to best utilize LLMs. 
% We relaxed the constraint to have an automated process that creates prompts, consistently generates high-quality suggestions, and times the LLM's response. 
% \am{stuff}
% Through this process, we can obtain high-fidelity observations of how analysts work conduct an analysis with an assistant. Analysts are also better able to provide specific feedback about the assistant and its suggestions. Moreover, a \woz approach helps us reliably ensure high-quality suggestions and control the types of suggestions analysts encounter.

\xhdr{Participants}
% With our study, we wanted to understand analysts' preferences towards and the subsequent impact of planning assistance. We imagine novice analysts may be too bogged down by execution challenges that there never is any time or consideration for analysis planning (even if offered planning assistance). Moreover, if even trained analysts struggle with analysis planning \cite{Kale2019DecisionMakingUU, Liu2019PathsEP}, then understanding how to support these analysts is a crucial first step to supporting all analysts. Therefore, we focused on recruiting data analysts who are already familiar with statistical analysis. 
We recruited participants who were already familiar with statistical analysis and programming experience.
% and were likely to engage in analytical reasoning.
We contacted potential participants through analysis-related mailing lists at our institution.
From a pool of 60 volunteers, we invited those who rated their programming and statistics proficiency 4 out of 5 or greater and were familiar with computational notebooks, selecting a final 13 on a first come basis (Table \ref{tab:participants}).
% We imagine \tim{can we cite related work to inform this expectation and provide some justification?} novice analysts may be too bogged down by execution challenges that there never is any time or consideration for analysis planning (even if offered planning assistance). 
We chose not to consider novices, as they would be too limited by execution challenges \cite{Liu2020UnderstandingTR, Kandel2012EnterpriseDA} to be able to benefit from our analysis planning.
% in its current form. 
% Similarly, we thought that individuals lacking general statistical knowledge would struggle to comprehend the full range of recommended decisions \cite{Liu2019PathsEP}. 
Since even trained analysts face difficulties in analysis planning \cite{Kale2019DecisionMakingUU}, supporting those with expertise is a crucial step in supporting everyone.
Participants received a \$50 gift card as compensation. We denote participants by anonymous identifiers, like AX, and quote them, \shortquote{like so}. 

\begin{table*}
    \small
 
  \begin{tabular}{lllllcc}
    \toprule
    \textbf{ID} & \textbf{Gender} & \textbf{Occupation/Background} & \rr{\textbf{AI Code Assistant Usage}} & \textbf{Prog Lang} & \textbf{Lang Exp} & \textbf{Stats Exp} \\
    \midrule
    A01 & Male & Professor in Public Policy  & None & R & 4 & 5 \\
    A02 & Male & PhD Student in Public Policy & None & R & 5 & 4 \\
    A03 & Male & PhD Student in Civil Engineering & None & Python & 5 & 4 \\
    A04 & Male & Research Scientist in Public Policy & Minimal & R & 5 & 4 \\
    A05 & Female & Masters Student in Data Science & Regular & Python & 4 & 4 \\
    A06 & Female & Masters Student in Data Science & None & Python & 4 & 4 \\
    A07 & Male & Masters Student in Data Science & Minimal & Python & 4 & 5 \\
    A08 & Female & PhD Student in Atmospheric Sciences & Minimal & Python & 5 & 5 \\
    A09 & Female & PhD Student in Political Science & None & R & 4 & 4 \\
    A10 & Male & Data Scientist & Regular & Python & 5 & 5 \\
    A11 & Female & Masters Student in Data Science & Regular & Python & 4 & 4 \\
    A12 & Male & PostDoc in Materials Science & None & R & 4 & 4 \\
    A13 & Female & PhD Student in Computational Finance & None & Python & 4 & 4 \\
    \bottomrule

  \end{tabular}
  \caption{\textbf{Participant Information.} \rr{Prior AI code assistant experience was gathered from our study interviews.}}
   \label{tab:participants}
\end{table*}

\xhdr{Procedure}
We conducted our study in a lab on a MacBook Pro on a 27-inch monitor. One author, in the lab with participants, served as the coordinator. Another author, the wizard, participated from a physically separate room over Zoom (which was also used to record the session with consent) under a pseudonym. Participants were unaware of the wizard and that the wizard was able to hear and see their screen.

\begin{figure*}[t]
  \centering
  \includegraphics[width=0.90\linewidth]{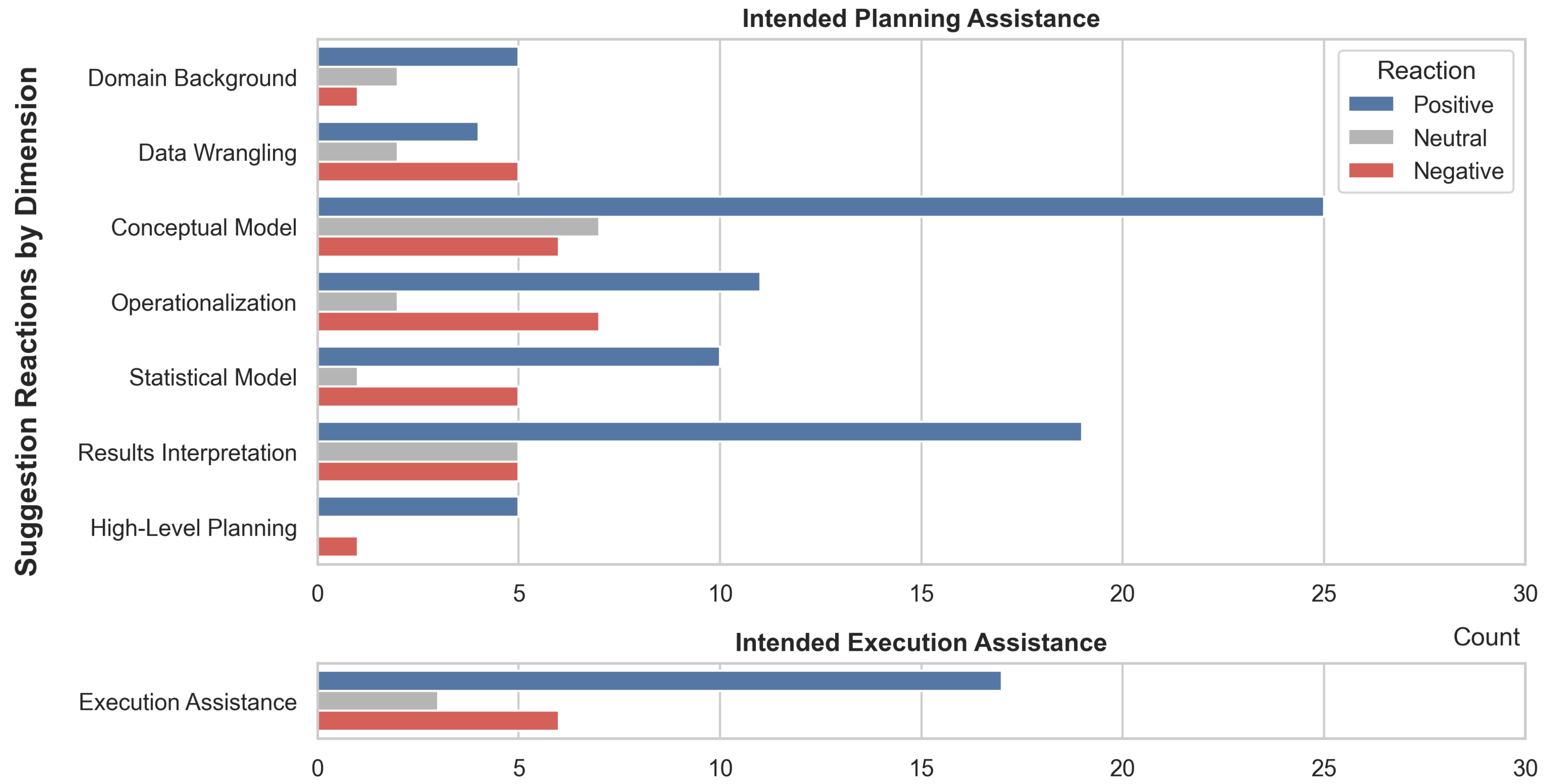}
  \caption{\textbf{Analysts favor currently unsupported planning assistance}. \rr{Analysts saw on average 11.85 (std=4.56) suggestions, 9.85 (std=4.16) of which were planning suggestions}. All categories of \textit{intended} planning suggestions (which may include code to help execute a suggestion) were found to be helpful by at least some analysts, highlighting the need for planning assistance that current assistants lack. \rr{For planning suggestions which analysts could reasonably incorporate into their notebooks (i.e., those that were not \textit{results interpretation} and \textit{domain background}), analysts integrated suggestions 51.6\% of the time (47/91).} Analyst reactions varied within each category, with responses ranging from finding the assistance helpful to neutral, or even unhelpful in certain instances. Our interviews and observations suggest that the effectiveness of a suggestion is not solely determined by its suggestion category but also depends on various nuanced factors~(Sec.~\ref{sec:results}). We provide full examples of the raised suggestions in the appendix.
  % \tim{I didn't quickly find where you refer to this primarily. can it be closer to that?}
  }
  \Description{Two vertically stacked horizontal bar charts. The top chart is titled "Intended Planning Assistance". The y-axis is each suggestion category and the x-axis is count. Each category (y-axis tick) has three bars representing "Positive", "Neutral", and "Negative" reactions to the suggestion category. The bottom chart is titled "Intended Execution Assistance" and has one category, Execution Assistance, on the y-axis. }
  \label{fig:suggestion_reactions}
\end{figure*}

The study lasted roughly 2 hours and consisted of three phases, consisting of a tutorial ($\sim$15 minutes), the primary task detailed below ($\sim$75 minutes), and a semi-structured interview ($\sim$30 minutes). 
In the last of these phases, the coordinator asked participants to review each suggestion they saw, give a positive, neutral, or negative reaction to whether the suggestion was helpful, and explain why (Fig.~\ref{fig:suggestion_reactions}). 
% \tim{do what degree did they do this live vs all retrospective during this interview?} 
We then asked about the impact the assistant's recommendations had on the participant's process, what participants valued from the assistant, and what sorts of actions or information they would have liked the assistant to provide. 
At the end of the study, we revealed how the assistant actually worked.
% , that is, its suggestions were a mix of manually written and ChatGPT-generated text, and its curation and raising of suggestions were done by a human wizard. 
Prior to the study, participants were asked about their preferred programming language and the analysis libraries. We created a notebook for each participant with their preferred language and tools. We wanted to ensure that the lab environment was as familiar as possible.
The study coordinator took notes throughout the analysis phase and the semi-structured interview. One author viewed the recordings, transcribed relevant episodes, and logged whether suggestions were included in participants' working analyses. To define common themes that emerged, two of the authors conducted iterative open coding on the recordings.

\begin{figure*}[t!]
  \centering
  \includegraphics[width=0.95\linewidth]{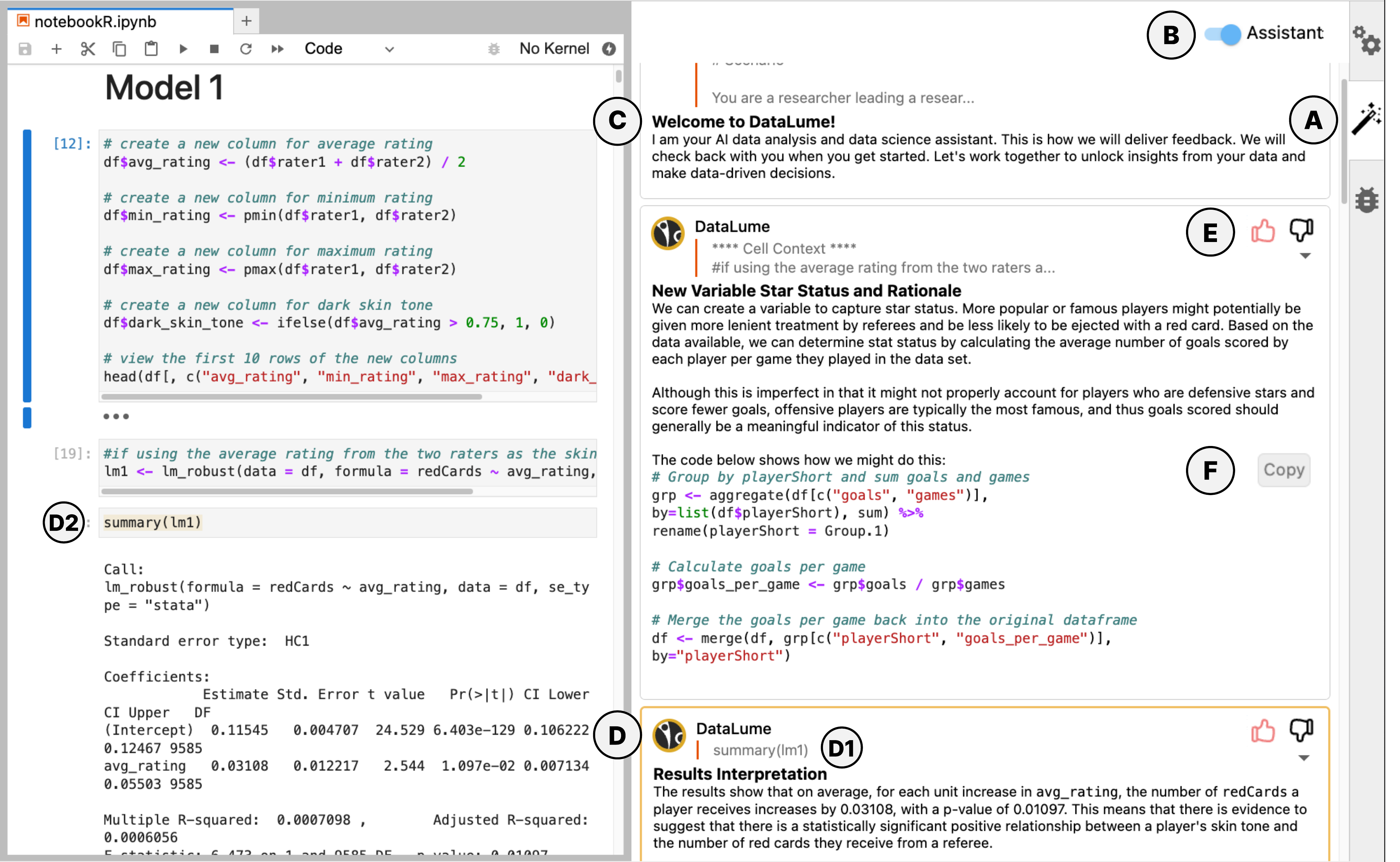}
  \caption{\textbf{Assistant User Interface}. Our assistant is integrated into JupyterLab as a side panel. Analysts can open the side panel in the right sidebar (A) and turn the assistant on/off with a toggle (B). Suggestions are raised as cards in this side panel. When first turned on, the assistant welcomes the analyst with a startup message (C). When a new suggestion occurs, the side panel shows a loading animation of variable duration to call attention to it; variability in duration aims to indicate that the system is working and simulate a real system computing and generating a suggestion~\cite{Fraser1991SimulatingSS, Riek2012WizardOO}. New suggestions also have a yellow highlight around them for greater visibility (D). Likewise, each suggestion has a corresponding context (D1) in the notebook, which is automatically scrolled into view when the suggestion is clicked (D2). Analysts can provide feedback for each suggestion via the thumbs-up and thumbs-down buttons (E). Finally, analysts can copy recommended code using the copy button (F). Note that the width of the side panel can be adjusted and is typically smaller than is pictured here for illustration.
  % \jeff{Figure has F and G, not E and F!}
  % \tim{Minor: match letters across figures. fig 1 uses colorful rectangles with rounded corners :)}
  }
  \Description{The figure shows a screenshot of the JupyterLab notebook interface with the notebook on the left side and the suggestion panel on the right side.}
  \label{fig:interface}
\end{figure*}

\xhdr{Task} The bulk of the study consisted of a data analysis  task---namely the task conducted in the crowd-sourced study\footnote{In the interest of time, we made some minor adjustments to the dataset.
We sampled a subset of the data to simplify computational manipulation while maintaining the overall distribution of the dependent variable (red cards) across levels of the main independent variable (skin tone).
Additionally, we focused on the ten most frequently used variables across analyst teams in the original crowd-sourced study.} considered in Sec. \ref{sec:suggestions}---in the context of a JupyterLab notebook customized to have a planning equipped assistant (Fig. \ref{fig:interface}).  
Participants were asked to imagine they were leading a research team that had collected the dataset, following a similar methodology from Jun et al.~\cite{Jun2021HypothesisFE}.
% \tim{I'm confused -- I thought someone other than Eunice collected the red cards data?!} 
To prime participants,
% to consider how their decisions could affect results, 
we told them that their analysis results would impact a major policy decision---namely whether the soccer league invests money in bias training.
We stressed the importance of having reasonable rationales for their analysis decisions and that their decisions must be robust against alternative assumptions.
Likewise, to encourage comfort with the lab environment, we mentioned that we were not interested in the completion of the analysis but in their analysis process.

Once participants were familiar with the task, the coordinator toggled on the assistant and introduced how its suggestions were raised. We described what the system can do (e.g., make suggestions based on the notebook context), being careful not to anthropomorphize the assistant to facilitate accurate evaluations~\cite{Khadpe2020ConceptualMI}. 
% \tim{curious. why is the latter important?}
We encouraged participants to think aloud or document their process in a notebook. They were free to use any external resources. 

\xhdr{Assistant Design}
The design of our assistant closely followed recent explorations for assistants in notebooks~\cite{McNutt2023OnTD} and is shown in Fig. \ref{fig:interface}. We focused on notebooks because they are an extremely common medium for data analysis~\cite{Shen2014InteractiveNS, Rule2018ExplorationAE}.
% \subsection{User Interface}
% \am{intrp?????}
% We chose to integrate our assistant in computational notebooks\footnote{We include details of the interface implementation in the appendix}, a popular medium for data analysis~\cite{Shen2014InteractiveNS, Rule2018ExplorationAE}. The design of assistant UI drew on McNutt et al.'s~\cite{McNutt2023OnTD} design space of code assistants in computational notebooks.
% through surveying existing notebook systems and performing an interview study. 
% \tim{this sounds like what you want  to do in this paper? can you add a couple of words to clearly differentiate their contributions from yours?}\ken{did this in related work, still necessary here?} 
% A central design decision was the location of the assistant within the notebook UI. We opted for a side panel interface, which offers a global perspective of access to information for the assistant. This global scope is helpful because it implies that the suggestions integrate information from all parts of an analyst’s notebook. Additionally, by keeping the suggestions separate from the analyst's working notebook, we minimize any interference with an analyst's code, notes, and overall process.  
We place our assistant within a side panel which affords an apparently global perspective across the notebook. This scope helpfully implies that the suggestions integrate information from across the notebook. By keeping the suggestions separate from the participant's working notebook, we minimize any interference with their code, notes, and overall process.  
As in other triggerless systems~\cite{jayagopal2022exploring}, participants received suggestions from the system without needing to take specific action. Given this design, there is no direct UI element for controlling the generated topics or to refine suggestions (similarly to Copilot's tab view). Also like Copilot, participants could influence the assistant by writing specific comments in their notebook. 
% While we did not not explicitly describe this interaction to participants, this behavior naturally emerged across multiple participants.
This comment writing behavior was not described in our introduction but emerged naturally across multiple participants. 
As suggestions from the wizard began to reference participant comments as the relevant notebook context, some participants took notice and wrote additional comments, expecting assistant help. 

\xhdr{Raising and Creating Suggestions} 
When and what suggestions the wizard raised were guided by several principles. 
The wizard raised suggestions when they were relevant to the notebook, with priority given to the most recent additions \cite{Amershi2019GuidelinesFH}. When possible, the wizard offered planning suggestions,
% , which are less well understood and supported by current systems (Sec. \ref{sec:llm_background}). 
% To maintain trust and show contextually relevant information \cite{Amershi2019GuidelinesFH, Horvitz1999PrinciplesOM}, 
but execution assistance was provided when it was deemed required. 
Examples include when participants were stuck debugging outputs or wrote notebook comments asking for specific help (Sec. \ref{sec:results_how}).
% \tim{were they instructed that they could do that? did it happen naturally? how many people did that?}
The wizard attempted to strike a balance between a rich diversity of suggestions and not overwhelming the participant with excessive feedback. 
Finally, we simulated an assistant that could learn over time what suggestions the analyst found helpful--following common Human-AI guidelines~\cite{Amershi2019GuidelinesFH}. To wit, the categories of suggestions we raised depended on whether the suggestions already given were considered or taken. If the participant disregarded many suggestions that addressed alternative variables, we raised these suggestions less often.

Creating suggestions followed the analysis conducted in the previous section (Sec.~\ref{sec:suggestions}).
For categories such as \textit{domain background} and \textit{operationalizing constructs} (Table \ref{tab:suggestions}), the same or very similar suggestions could be given to different participants. To speed up the assistant's response time, we prepared a spreadsheet of suggestions (based on Table~\ref{tab:suggestions}) in these categories before the study for the wizard to draw from (Fig. \ref{fig:woz}B). 
% The spreadsheet had the categories in Table \ref{tab:suggestions}, except each category could have multiple prepared suggestions. 
We created 32 suggestions across these categories. We deliberately crafted all suggestions with pertinent explanations and relevant domain and statistical knowledge. For most suggestions, we also included example execution assistance code to help participants grasp and implement the given recommendation. Though these planning suggestions featured execution assistance elements, their primary intent was to aid in analysis planning.

For categories \textit{execution assistance} or \textit{model results interpretation} (Table \ref{tab:suggestions}), suggestions were closely tied to the context of the working analysis which precluded previously created suggestions.
% Therefore, suggestions created previously could not be used. 
For these categories, we leveraged ChatGPT \cite{ChatGPT} and constructed prompts to build suggestions. Prior to the study, we created a general prompt introducing the task and dataset.
% and specific prompts for different categories. 
% During the study, ChatGPT was tightly integrated into the suggestion creation process. 
% \tim{we should be consistent. are we building/developing/creating suggestions? I don't have a strong opinion but like creating.} 
In addition to this general prompt, the wizard, who prepared the spreadsheet of suggestions and is experienced in using LLMs, crafted prompts in real-time using the notebook context (Fig. \ref{fig:woz}A) or parts of pre-written suggestions (Fig. \ref{fig:woz}B). The wizard then passed the prompt to ChatGPT to generate the suggestion of interest (Fig. \ref{fig:woz}C). 
% Given their experience and preparation, 
% the wizard primarily wrote prompts in real-time. 
% \tim{didd you do these from scratch every time? it reads that way.}\ken{basically yes because it was so varied with approaches and prompts were not super difficult to write in the momment} \tim{okay. then say that the wizard did write these prompts largely in real-time, but their experience allowed them to do this relatively quickly} 
% Further, the wizard used 
ChatGPT was further used to refine some generated code---for example, to convert code between programming languages or shorten the length of a prior generation (Fig. \ref{fig:woz}C-D loop). 
The wizard sometimes manually corrected the output of a ChatGPT generation (e.g., changed the wording or variable names) to ensure high-quality suggestions. 

\xhdr{Limitations} While able to elicit a variety of realistic user reactions to a new form of assistant this design has some limitations.
While a complete implementation of full system would be useful, our goals in this work are to better understand the design of such assistants rather than their implementation. 
Similarly, through this approach we are not limited by the capabilities of current LLMs (Sec. \ref{sec:llm_background}) or participants non-expert prompt engineering abilities~\cite{ZamfirescuPereira2023WhyJC,  Mishra2022HELPMT}.
% \am{sentence}
Having a wizard create and deliver suggestions also raises a potential downside of being slower to respond than an automated system. However, we chose to conduct our study with a single analysis task and dataset so we could prepare high-quality suggestions in advance. 
% In particular, we used the same task and dataset as the one from the crowd-sourced analysis study we examined~(Sec. \ref{subsec:crowdsourced_study}). 
This helped us both ensure a high quality and rich diversity of suggestion content and timing while being only slightly slower than a fully automated approach. 

%%%% 5-woz-new.tex ends here %%%%

% \input{5-woz}

% \input{6-study}

%%%% 7-results.tex starts here %%%%

\section{Results}
\label{sec:results}

\begin{figure*}[t]
  \centering
  \includegraphics[width=0.99\linewidth]{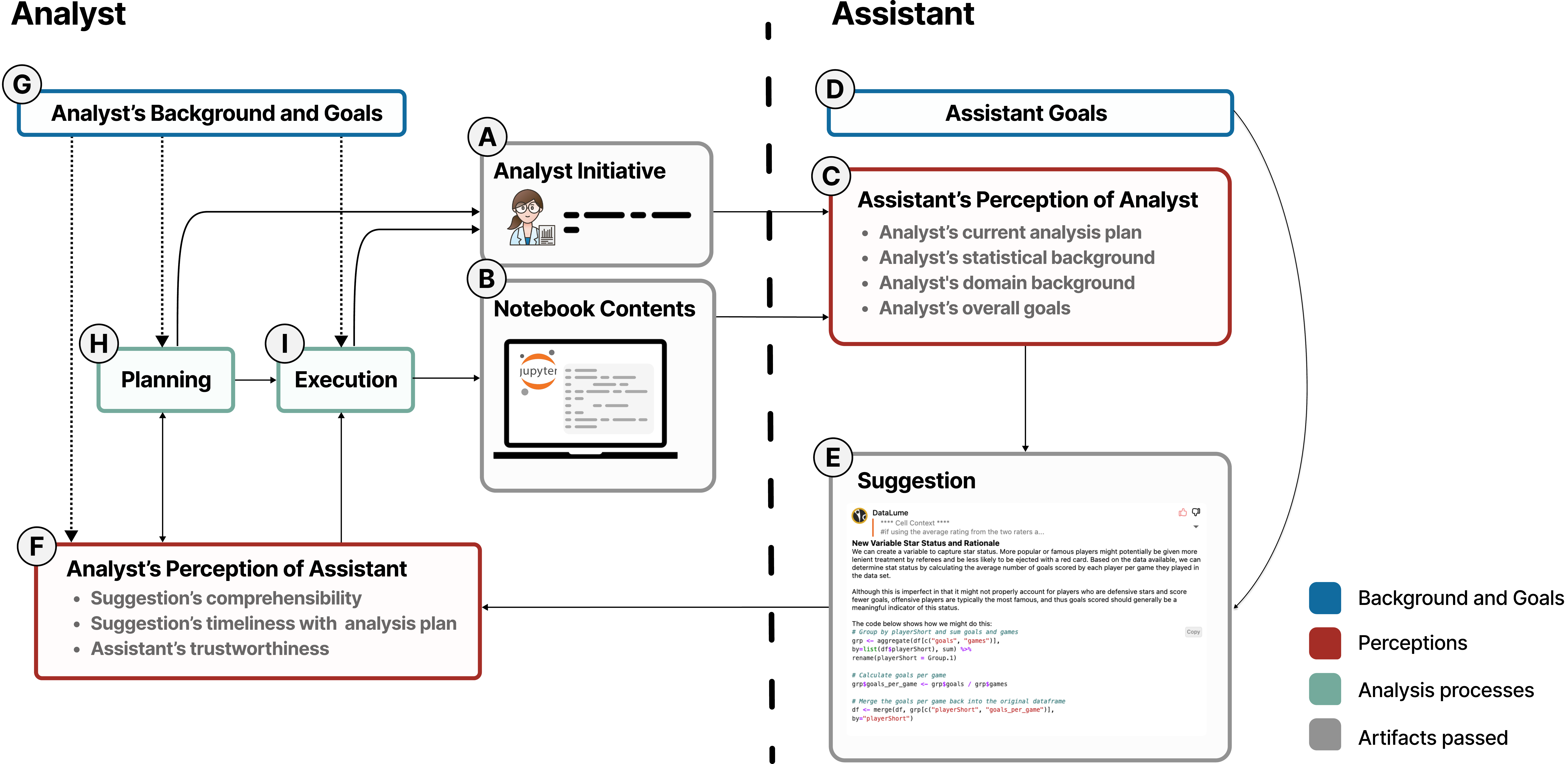}
  \caption{\textbf{Study-informed Model of Analyst-Assistant Interaction Dynamics}. Based on our study findings, we model the underlying influences and interactions between an analyst and an assistant. The assistant can receive information about the analyst's initiative (A) and the analysis contents (B) to develop an understanding of the analyst's goals and background as well as the current analysis plan and context (C). Note that the analyst's initiative is optional. This, in conjunction with the assistant's programmed goals (D) (e.g., offering more planning assistance or doing only what analysts request), informs the assistance (i.e., timing, organization, location, execution vs. planning, and contents) provided to analysts (E). After a suggestion is raised to the analyst, how it is perceived (F) is impacted by the analyst's background (G) and their current analysis plan (H). This then determines the extent to which the analyst accepts the suggestion in their planning (H) and/or execution (I) processes. The loop then continues as analysts update their analysis (B) or require additional assistance (A).
  % \tim{since this come sat the top of results I think it's possible to mistake this figure for your how system/wizard worked (e.g. ``programmed goals''. I think you should say clearly in the bold text and first caption sentence that this is a model of interaction dynamics that is informed by the results of your study.}
  % \jeff{I don't totally follow the ordering here. For example the caption mentions G and H before F.} fixed ordering of sentence
  }
  \Description{A diagram split down the middle vertically separating the components on the analyst side on the left and the assistant side on the right. The analysts side includes the following components written in boxes: analyst's background and goals, analyst initiative, notebook contents, planning analysis process, execution analysis process, and analyst's perception of the assistant. The assistant side includes the assistant goals, assistant's perception of the analyst, and suggestion. Arrows point between these components to inform a  model of analyst-assistant interaction dynamics.}
  \label{fig:results}
\end{figure*}

% From our \woz study, we observe \tim{be consistent in terms of present  or past tense} many interesting behaviors and reactions to our system that bring new insights into  Human-AI collaboration in the context of a data analysis assistant. 

Our \woz study explored two main questions: ``\emph{what characterizes helpful suggestions?} (\textbf{RQ2}), and \emph{how do these suggestions impact analysts' workflows?} (\textbf{RQ3}).

With respect to \textbf{RQ2}, we found that analysts preferred suggestions that matched their analysis plan and their statistical and domain background (Sec.~\ref{sec:results_who}).  They preferred both execution \textit{and} planning assistance, though planning required more cognitive effort (Sec.~\ref{sec:results_what_assistance}). As a result, explanations provided in code, code comments, and natural language greatly assisted analysts in understanding the intent and rationale of suggestions (Sec.~\ref{sec:results_what_qualities}). Due to the extra effort to consider planning assistance, analysts valued suggestions that were well-timed to their current task (Sec.~\ref{sec:results_when}).

For \textbf{RQ3}, we found well-timed and contextual planning assistance often helped analysts consider and make alternative decisions (Secs.~\ref{sec:results_what} and ~\ref{sec:results_when}). In contrast, analysts were reluctant to accept suggestions that mismatched with their expertise or task. Likewise, we observed potential drawbacks of AI-assistance: some analysts became distracted or overly reliant on the assistant's suggestions, often without taking the time to critically assess their relevance or correctness (Sec.~\ref{sec:results_why}). Nevertheless, these impacts were not uniform across all scenarios and varied depending on the individual analyst (Sec.~\ref{sec:results_who}), the type of assistance (Sec.~\ref{sec:results_what}), and when the suggestion was made (Sec.~\ref{sec:results_when}).

% Overall, analysts' reactions to raised suggestions  show that suggestions from all categories were helpful, 
Overall, analysts generally saw suggestions from all categories as helpful (Fig. \ref{fig:suggestion_reactions}),
% although some found the assistance to be unhelpful in certain situations (Sec.~\ref{sec:results_who} and Sec.~\ref{sec:results_when}). 
some situations excepted (Secs.~\ref{sec:results_who} and ~\ref{sec:results_when}).
Triangulating their ratings with interview data revealed subtleties in analyst-assistant interaction dynamics, which we summarize in Fig. \ref{fig:results}. 
 Here, we consider suggestion helpfulness, their analysis impact, and the overall experience using a "\emph{Who, What, When, Why, and How}" framework to organize our findings.
While analysts expressed preferences for ``\emph{where}'' the assistant should be placed on the UI, these preferences did not fundamentally impact their overall workflow, and so we include our observation to this effect in the appendix. 

\subsection{\emph{Who}---Analyst Background}
\label{sec:results_who}

Our study revealed many aspects of an analyst's background that influenced their perception of suggestions and overall experience. Here, we discuss their analysis plan (Sec. \ref{sec:results_who_analysisplan}), statistical and domain backgrounds (Sec. \ref{sec:results_who_expertise}), and prior experiences with and curiosity about AI-based assistants (Sec. \ref{sec:results_who_curiosity}).

\subsubsection{Analysis Plan}
\label{sec:results_who_analysisplan}
Analysts exhibited varying levels of analysis forethought and rigidity in their analysis plans. While all analysts had a rough plan, some took extra time to explicitly detail their analysis steps [A3, A5, A9, A10, A11]. For example, A9 spent the first 20 minutes planning on scratch paper before writing a single line of code. 
Others employed on familiar strategies, for instance, A4's initial plan drew on their familiarity with  Species Distribution Modeling \cite{Franklin2010MappingSD}.
% ---a statistical approach that predicts species' geographic distribution based on environmental variables.
% In other instances, analysts considered approaches that were familiar. F
% for instance, A4, who was familiar with Species Distribution Modeling \cite{Franklin2010MappingSD} (i.e., a statistical approach that predicts species' geographic distribution based on environmental variables), mentioned that approach as an initial idea for planning. 
% \tim{sounds interesting -- was this at all related to the task?}
% Likewise, 
Analysts (8/13) often liked suggestions that matched their own ideas or plans.
% [A3, A4, A5, A7, A9, A11, A12, A13]. 
They appreciated that the included code offered additional execution assistance to implement their plan. 
For example, in reference to a suggestion recommending Age as a variable, A12 noted \shortquote{I really liked this one... I think it actually somehow knew that I was going to do this, and then it suggested a very efficient way to do so.} 
Furthermore, for A5, such suggestions reaffirmed their decisions: \shortquote{The assistant actually gave the same conclusion that I had typed, so that was helpful.} 
% Overall, analysts frequently favored suggestions that matched their analysis plan. 
% \am{weak return}
% Given variations in how they plan, they also wanted ways to specify the level of assistance they received~(Sec.~\ref{sec:results_how}).
Given these planning variations, it is unsurprising that participants also expressed interest in controlling the level of assistance received~(Sec.~\ref{sec:results_how}).

\subsubsection{Statistical and Domain Knowledge Expertise}
\label{sec:results_who_expertise}
% We observed disparities in individual levels of statistical and domain expertise across analysts and throughout the course of the analysis. This affected how analysts reacted to the assistant's suggestions. 
Analysts' varying levels of statistical and domain expertise affected how they reacted to presented suggestions.
Most analysts (11/13) had no trouble understanding the feedback for most suggestions, with many (10/13) finding them to be well-matched with their expertise and helped them consider new approaches.
% [A1, A2, A3, A4, A5, A7, A9, A10, A11, A12, A13] \jeff{This is a rather long list, maybe (11/13 participants) or somesuch would be more useful?} \tim{+1 (and you can keep the records e.g. in comment or elsewhere}. 
% Likewise, 10/13 analysts found suggestions helpful when they were well-matched with their expertise and helped them consider new approaches.
% [A2, A3, A4, A5, A6, A7, A9, A10, A12, A13]. 
% For example, A7 appreciated an important reminder from the assistant about how to encode a categorical variable: \shortquote{This is a great suggestion. I didn't know I could just do that instead of having to encode it. So it takes care of it instantly.} However, a few analysts found some suggestions overly basic for their own expertise and therefore did not benefit from them [A1 A3, A5, A7, A8]. For example, A1 did not gain any insight from a results interpretation suggestion: \shortquote{It was giving me comments on my results. It was not telling me whether it was good or bad which I knew already.} However, these same analysts also noted that the suggestions would be quite helpful for beginners [A1, A3, A7].  
However, some analysts found suggestions overly basic and unnecessary [A1 A3, A5, A7, A8], due to their statistical expertise. This was most evident with the results interpretation suggestions created using ChatGPT. Some analysts liked these suggestions [A5, A6, A7, A10, A11, A12], while others adopted a more neutral stance [A8, A4]. A1, a statistics professor with extensive expertise in computational social science, regarded the suggestions as unhelpful: \shortquote{It was giving me comments on my results. It was not telling me whether it was good or bad which I knew already.} 

Suggestions were occasionally unfamiliar since some analysts lacked an adequate statistical background [A3, A5, A6, A8, A9, A13]. This led to varying behaviors: A3, A5, and A13 chose to ignore unfamiliar suggestions completely;
A6 and A8 spent time trying to understand the suggestion but were left confused; and A9 considered referencing external sources to interpret the suggestions: \shortquote{My first idea is to try and copy and search on the internet to see how this kind of generalized linear model could help me.}
Suggestions that were mismatched with the analyst's statistical background were unhelpful or required significant additional effort to understand, highlighting that assistant usage might usefully begin with an alignment phase. 
% This compounds the already intense concentration and effort needed to consider planning assistance~(Sec.~\ref{sec:results_what_assistance}).\am{suggesting what}

% \subsubsection{Prior Experience with and Curiosity about AI-based Assistants}
\subsubsection{Prior Experience with AI-based Assistants}
\label{sec:results_who_curiosity}

We found that analysts' own past experiences with AI led to behaviors and reactions that diverged from the immediate analysis task. 
Analysts had mixed prior experiences working with AI-based code assistants (i.e., Github Copilot, ChatGPT). Over half (7/13) had no experience while others ranged from using code assistants a few times in total to regularly using them (Table~\ref{tab:participants}). 
These prior experiences caused some analysts to be skeptical about suggestion correctness. For instance, A8 and A13 questioned the resources that informed it.
Referring to a suggestion that referenced external studies, A8 mentioned, \shortquote{I have less trust of something that is a more generated summary of stuff that exists outside of the notebook, that I didn't know where the information came from.}
% \tim{do you discuss this later? interesting and important observation/design rec around trust}
% In contrast, A4 was surprised by the suggestions the assistant provided given prior experience using ChatGPT: \longquote{I am very surprised because when you ask ChatGPT about something really obscure, it might give you good information or something bullshit, and this seemed tightly reasoned. It's a little spooky.}\am{comment on the wizard of oz study nature}
Some analysts became distracted by the novelty of using an assistant. For instance, A8 was distracted from their task by their curiosity: \longquote{It made me want to understand how this (worked). I was getting distracted. What is it learning from? Is it from the data and knows all the variables or from what I am asking for? I feel like I was experimenting a lot with putting little notes and seeing what would happen.}  
% We suggest that as AI-based assistants become ubiquitous, such issues will likely be more limited.
% Without knowing many details about how the system worked, some analysts were curious about the assistant [A1, A4, A8, A12] \am{in what capacity}. 
% % A4, for example, was open to what the assistant suggests: \shortquote{We can try a basic linear model, which should not be very good, but I'm curious to see what (the assistant) tells me after that.} 
% % A8, on the other hand, became more distracted by their curiosity: 
% For instance, A8 was distracted from their task by their curiosity: \longquote{It made me want to understand how this (worked). I was getting distracted. What is it learning from? Is it from the data and knows all the variables or from what I am asking for? I feel like I was experimenting a lot with putting little notes and seeing what would happen.}  
% We suggest that as AI-based assistants become ubiquitous, such issues will likely be more limited. 
Our observations agree with existing work~\cite{Liao2022DesigningFR} that models trustworthy AI, where a user's attitude towards an AI system (and therefore subsequent reactions to the system) depends on individual, environmental, and cultural contexts in addition to the system's trustworthy cues.
% \mgm{Conclusions to each section may be helpful, esp if people are skimming. What was agreed upon vs not among different analysts}

\subsection{\emph{What}---Suggestion Content}
\label{sec:results_what}
% We noticed broader qualities relating to \jeff{Sentence incomplete!}
Though we found that the exact category of planning content (e.g. Table \ref{tab:suggestions}'s categories) was less important for depicting the helpfulness of a suggestion~(Fig.~\ref{fig:suggestion_reactions}), the type (planning and execution) and specificity of assistance were preferred in different situations (Sec. \ref{sec:results_what_assistance}). Moreover, we found well-reasoned explanations to be important to analysts (Sec. \ref{sec:results_what_qualities}).

\subsubsection{Execution vs. Planning Assistance}
\label{sec:results_what_assistance}
All analysts appreciated execution assistance, occasionally rating it negatively only when it came too late or differed from what they had intended. In our study, execution assistance occurred either on its own---when analysts were struggling with executing their plan---or as part of intended planning suggestions. 
Analysts (8/13) appreciated that execution assistance saved time and prevented them from having to search the internet.
% [A1, A3, A4, A5, A7, A9, A11, A12]. 
For example, A3 liked that a suggestion to consider a BMI variable included the code to calculate it, noting that \shortquote{I don't know how to calculate BMI off the top of my head, and it was like `Hey, here's how you calculate it'. Perfect, saves me a Google.} These preferences are in line with prior observations~\cite{Vaithilingam2022ExpectationVE, Ziegler2022ProductivityAO} that expediency and reduced online searching are major perceived benefits of AI-based code assistants. 
Some analysts would have preferred even more execution assistance [A2, A3, A4, A5, A13]. For example, A5 and A13 expected more help when faced with errors, while A3 wanted it to make visualization for them.
% . Likewise, A3 wanted the assistant to execute the code for them, such as for visualizations.  
% \mgm{i think there was more people who wanted this, on the Mural from day 1 --> my side of stickies --> interaction considerations --> prefers automatic, it was P4 too and then 2 other stickies that don't have a Participant written down}. 
% Similarly, A2 wanted high-level help with \shortquote{understanding the differences between libraries.}

Reciprocally, most analysis found planning assistants to be helpful, although this utility 
% With respect to planning assistance, most analysts found it to be helpful. Helpfulness 
depended less on suggestion type than on the timing (Sec. \ref{sec:results_when}).
% with respect to analysts' own paths (Sec. \ref{sec:results_when}). 
Most analysts (12/13) noted that planning suggestions often presented ideas that they had not previously considered.
% [A2, A3, A4, A5, A6, A7, A8, A9, A10, A11, A12, A13]. 
For instance, A10 contrasted our assistant's planning merits with other assistants: \longquote{Auto-complete code (assistants) are very good, but they are looking at what you are doing right now and maybe the last cell. This agent could take a more overall approach and help you think about the overall approach.}
In other situations, the suggestions also helped analysts realize key steps they would have otherwise missed. For instance, A2 found observed that  \shortquote{this is helpful. I didn't realize that skin tone rating was a combination of rater one and rater two.} 
% \shortquote{Interesting, this is helpful. I didn't realize that skin tone rating was a combination of rater one and rater two.} 

However, while most suggestions the assistant presented were tied to specific moments in the analysis process, some analysts wanted broader analysis planning help [A2, A3, A6, A7, A8].  A3, for example, mentioned how at the beginning of the analysis, \shortquote{recommendations for how to think about the problem would (have been) good.} Similarly, A2 wanted suggestions that could provide a whole analysis plan: \longquote{Given the dependent variables and the data structures, what are some of the analyses I could run? If I need to run different analyses, how would I need to transform the data, with suggestions or sample code of how I can do that?} In these situations, analysts might benefit from workflow scaffolding recommendations similar to those in \textit{Lodestar}~\cite{Raghunandan2022LodestarSI} or \textit{litvis}~\cite{Wood2019DesignEW}.

\subsubsection{Suggestion Explanations}
\label{sec:results_what_qualities}
Several analysts  [A2, A5, A6, A7, A8, A9] specifically noted that they liked the explanations included in the suggestions. 

For A5 and A9, both the included code and comments acted as explanations that helped them understand what the assistant was trying to do. 
% \tim{bit confused here about "explanation" vs "code". is the comments the explanation?} 
A9 felt that explanations helped bridge a lack of trust in the assistant about a new calls-per-game variable suggestion, noting that \shortquote{I'm not trust(ing) with the calls-per-game, so I checked with every line of the code if they actually write what I'm thinking about.} Meanwhile, explanations for why a planning suggestion should be considered were also preferred. A2 was especially enthusiastic: \shortquote{I love the justification of why you would want to use age.} 
% Similarly, A7 valued the pros and cons offered by the explanation, which they felt gave them control of the final decision. 
In contrast, A8 connected their statistical expertise (Sec. \ref{sec:results_who_expertise}) to their willingness to accept a suggestion:
% they liked the suggestion overall and the explanation seemed to be logically sound, but they were not sure whether it was the right path to follow:
\longquote{I don't know if I was being led astray by the assistant but I found this very convincing. But I don't have the fresh statistical knowledge to actually know if it was a good statistical approach or not.}

The explanations were also seen as detrimental at times [A2, A6, A8, A13], depending on the analyst's expertise (Sec. \ref{sec:results_who_expertise}) and focus on their own analysis plan (Sec. \ref{sec:results_when}). A13 ignored suggestions when \shortquote{there is too much explanation}. Similarly, A2, who appreciated the explanation for the inclusion of age, thought that some suggestions contained too much text. 
% From a cost-benefit perspective~\cite{vasconcelos2023explanations}, the cost of engaging with the suggestions was too high.
A6 noted that \shortquote{Because I didn't understand why it was trying to include goals per game, I was kind of confused}. 
This is akin to how programmers working with AI-based code assistants sometimes struggle to understand the suggested code~\cite{Vaithilingam2022ExpectationVE}, highlighting the importance and fragility of explanation design.
% This sentiment is similar to those expressed by programmers working with AI-based code assistants; they struggled to understand the suggested code which hindered their ability to solve tasks ~\cite{Vaithilingam2022ExpectationVE}.
% Moreover, for A6, the explanation was insufficient and so confused them: \shortquote{Because I didn't understand why it was trying to include goals per game, I was kind of confused}. This sentiment is similar to those expressed by programmers working with AI-based code assistants; they struggled to understand the suggested code which hindered their ability to solve tasks ~\cite{Vaithilingam2022ExpectationVE}.
% \mgm{New paragraph? since not about explanations?} 

\subsection{\emph{When}---Timing of Suggestion}
\label{sec:results_when}
% A common factor that influenced analysts' perception of suggestions' helpfulness was the timing of suggestion presentation relative to their current progression and goals.
% The timing of when a suggestion was presented (relative to the current task) was central to whether or not analysts perceived it as helpful.
Timing has long been a central part of designing interactions with autonomous agents, and like previous studies~\cite{Horvitz1999PrinciplesOM, Fogarty2005PredictingHI}, we find that inopportunely offered suggestions risk being ignored.
% and \am{we find blah blah blah}
% in the end-to-end analysis process.  
% Many analysts found some suggestions to be irrelevant to their plan and thus unhelpful (9/13 analysts).
% [A1, A2, A3, A4, A5, A8, A10, A12, A13]. 

We observed that when engrossed in a particular analytical task, analysts were often resistant to diverting their focus toward distracting suggestions. 
% When analysts  [A5, A6, A10, A11, A12, A13] found the suggestions out of sync with their current analysis plans, they would ignore them to continue their previous line of thinking. 
For instance, A5 found that suggestions were often discordant with their current objectives:
% For A5, suggestions often seemed unresponsive to their current workflow: 
\longquote{very often the suggestions aren't in line with what you're currently typing. It goes in a whole different direction, and now you have to think of whether you want to take that direction or your direction.} 
Similarly, A3 wanted to focus on their current step: \shortquote{``I ignored it because I was focused on the variables at the time.} For A5, poor timing also led them to ignore it: \shortquote{I don't know if this was relevant or not. I didn't read it completely because I was trying to complete what I already started.} 
% Timing issues also slowed down A5's process since they had to stop and consider the merits of the suggested direction. 
% Given the costs of considering planning assistance, appropriate or non-intrusive timing of suggestions is even more crucial. 
Notably, most analysts (9/13) felt that ``unhelpful'' suggestions, would have been useful if they had been offered at a different moment.

% % In most situations, when analysts focused on their own plan, suggestions considered unhelpful at the moment would have been helpful if offered at a different time (9/13 analysts).
% % \longquote{(With) this assistant, very often the suggestions aren't in line with what you're currently typing. It goes in a whole different direction, and now you have to think of whether you want to take that direction or your direction.} In most situations, when analysts focused on their own plan, suggestions considered unhelpful at the moment would have been helpful if offered at a different time (9/13 analysts).
% % [A2, A3, A6, A7, A8, A9, A10, A11, A13]. 
% % Many analysts thought unhelpful suggestions would have been helpful later on [A6, A7, A8, A9, A10, A11].  
% % For A7 and A10, 
% % Yet, some suggestions that were unhelpful in the moment aided in future decision making. 
% For example, A7 mentioned: \shortquote{If I think it deviates too much, then I ignore it. There were a couple of times for the CHI-squared where I was like `Oh yeah that's a good idea.' I didn't do it at the time, but I put a pin on it.} In other cases, analysts  [A3, A6, A13] thought the suggestions came past after they were needed in their analysis. A6 would have preferred a covariate suggestion to come earlier: \shortquote{It did explain how it would be necessary, but maybe if it came before I did the covariate analysis (and) before I coded it out}.

When suggestions deviated only slightly from the current tasks analysts [A2, A5, A7, A9, A12] were open to them. For instance, while working through some statistical models, A7 took a recommendation to use the existing birthday column to create an age variable: \longquote{Honestly, birthday didn't hit me at that moment before the suggestion that popped up to calculate age. Without the suggestion, I wouldn't have thought of it unless I ended up reading about it and then including it a lot later.} This suggests that while suggestions should be aware of the current task, they need not \emph{exclusively} service that task as adjacent ones might usefully highlight alternative analysis paths---a key planning assistant goal.

% \tim{somewhat of a note to myself: do we discuss anywhere that "perceived helpful by analyst" may not be the only important outcome. They may all feel disrupted but what if it was an important thing they missed or that would have impacted their conclusions?}

Inherent in these observations is the tension between not disrupting the analyst and highlighting choices that might impact the analyst's conclusions. 
% Given the consequences of spurious analysis conclusions, the perceived helpfulness of suggestions may not be the only important outcome. 
% Similarly, 
From a cost-benefit perspective \cite{vasconcelos2023explanations}, execution assistance is less intrusive, with its time-saving benefits more easily recognizable. Conversely, planning assistance demands more mental effort as analysts evaluate multiple analytical paths, underlying rationales, and the alignment with their prevailing strategies. 
% \tim{and in addition implementing these plans likely would come with additional cost, right?}\ken{yes but this can be reduced through exec assistance. The ones mentioned so far are I think the most central at being costly} 
This complicates the assessment of the utility of planning assistance and amplifies the drawbacks of sub-optimal suggestions.
% We discuss this in Sec.~\ref{sec:discussion_helpful} and ~\ref{sec:discussion_balance}.

% Further, the perceived "unhelpfulness" of some planning suggestions was often due to \textit{when} these suggestions were presented~(Sec.~\ref{sec:results_when}).

\subsection{\emph{Why}---Reasons to Work with the Assistant}
\label{sec:results_why}
% \mgm{An intro sentence about what we mean by 'why' could be helpful, since first few sentences now summarize things that were already said}
% Analysts expressed a variety of preferences for \textit{why} they would want to work with a planning enabled assistant, inclyding its  its impacts on their workflows (\textbf{RQ3})
% We now describe analysts' overall experiences with the assistant and explore the reasons \textit{why} they want to use one. 
% These reasons are influenced by its impacts on their workflows (\textbf{RQ3}), which is in turn shaped by various elements of the assistant, such as the timing (Sec. \ref{sec:results_when}) and content (Sec. \ref{sec:results_what}) of the suggestions.

Most analysts (11/13) thought the assistant was helpful for both planning and execution-centric tasks, and expressed a variety of preferences as to \emph{why} they would want to work with this style of assistant.
% in their analysis process.
% [A2, A3, A4, A5, A6, A7, A8, A9, A10, A12, A13]. 
% They frequently liked the code that was provided (Sec. \ref{sec:results_what_assistance}), both for execution \textit{and} planning assistance. 
% For instance, A4 appreciated that the assistant saved them time by doing frequent coding tasks: \shortquote{The one thing that I really liked about it is that it saves me from having to do a lot of mundane stuff like setting up a testing and training dataset for some sort of machine learning model.} 

% \am{boiler plate and blank page problem are great make them be more featured}

%  (Sec. \ref{sec:results_what_assistance})
Analysts valued the planning assistance content provided by the assistant.
For instance, A9 found it helped them to consider alternative decisions: \shortquote{The (assistant) is helpful for me to indicate some aspects to think about. Though I did not take all of them, I see this is something that I may need to take a look at.} Furthermore, A2 liked having a clear rationale for why they should incorporate a suggestion (Sec. \ref{sec:results_what_qualities}): 
\longquote{As compared to when I'm typing in Gmail where it (suggests) words, this I find is much more helpful because it not only gives me the code but also gives the justification and rationale.} 
This suggests that planning assistance can be perceived as being of use to the data analysis process. 
% \longquote{I see a lot of potential and utility for this. As compared to when I'm typing in Gmail where it (suggests) words, this I find is much more helpful because it not only gives me the code but also gives the justification and rationale.} 

% Gordon \etal{} \cite{gordon2023co} notes that some of these beliefs 
% \am{this also has some to do with positioning and the culturallly understood role that it has, cf andy's ai-audit paper}

Even though our design primarily was centered on helping analysts consider alternative but reasonable approaches, they also frequently liked execution assistance. 
% Despite our design that mainly provided suggestions to help analysts consider alternative but reasonable approaches, analysts frequently liked execution assistance. 
For instance, A4 appreciated that \shortquote{it saves me from having to do a lot of mundane stuff like setting up a testing and training dataset for some sort of machine learning model.}
Similarly, A5 appreciated execution assistance for helping them move past roadblocks, noting that when \shortquote{you are stuck... and (the assistant saved) you that trouble of going somewhere else and googling things.}
% \shortquote{It was fascinating, to be honest. Because very often you're doing things, and you are stuck... and (the assistant saved) you that trouble of going somewhere else and googling things.}
This is consistent with the preferences found in prior works that studied code execution assistants \cite{Vaswani2017AttentionIA, Liang2023UnderstandingTU, Ross2023ThePA}, and suggests that having both both planning \emph{and} execution is critical for the design of this style of assistant. 
 
However, given the quality and quantity of assistance provided by our assistant, A8 believed that it made them less engaged in critical thinking. For A8, the reason to use the assistant was to reduce their own cognitive load rather than to have an AI guide their decision making: \shortquote{I feel like this `friend' was great, but it also made me this clicking machine (rather) than an analytical thinker... Autopilot was a welcome path that I could choose.} This follows the findings in empirical Human-AI collaboration studies in which people often over-rely on AI support~\cite{bansal2021does, buccinca2021trust, lai2019human, Buccinca2020ProxyTA}.
% am: todo, if accepted add cites to tarot paper + leilani and michael's health rec paper which also says this
% \am{wip} Similarly, the ``position''~\cite{gordon2023co} that the system tool (assistant rather than evaluator) was likely central to this perspective. 

% \tim{We should discuss this later if we haven't yet. danger to make things worse, especially if the assistant is not worthy of the trust put in it.}

% Similarly,  A4 appreciated the reasoning behind suggestions and compared it to that of existing assistants: \shortquote{I am very surprised because when you ask ChatGPT about something really obscure, it might give you good information or something bullshit and this seemed tightly reasoned.}

% \mgm{Everything here seems related to past sections (two of the quotes I had suggested as additions to prior sections lol). Might be worth discussing how to pitch this one as different. Like the effects of what the suggestion says and when / how it is raised depend on why the analyst considers the assistant helpful?}

\subsection{\emph{How}---Analyst or Assistant Initiative}
\label{sec:results_how}
% \mgm{defining initiative coßuld be helpful.}
% Finally, our interviews elicited comments about whether analysts preferred the assistant or themselves to initiate suggestions. 
Analysts espoused a variety of preferences about whether they wanted to initiate suggestions themselves or have the assistant do it for them. 
A few analysts [A4, A7, A9, A11] appreciated the assistant taking the initiative. For example, A11 liked that it \shortquote{proactively goes ahead and gives you suggestions.} 

Some analysts  [A2, A3, A4, A9] wanted the assistant to take even more initiative. A2, for instance, wanted the assistant to automatically run code given their current workflow: \longquote{[I would want the assistant to get...]  a sense of what my workflow is, has been, and is intended to be and pre-populate stuff for me. So if I normally run an OLS with varying degrees of variables, some included and some not included, generate a histogram for each. Do that for me.}

However, others felt that assistance would be more effective if it was delivered on demand.  
% \am{this repeats with that run visualizations for me from above}
Given the preference for both assistant and analyst-initiated assistance, 
multiple analysts [A6, A7, A9, A13] suggested that this could be a configuration option. 
% multiple analysts [A6, A7, A9, A13] suggested that this choice could be left for the analyst to control. 
To wit, A9 proposed: \longquote{Maybe two modes: lazy mode and (control) mode. If I am in lazy mode, you should already construct things and add them (to the notebook). If I want to take over everything, please be quiet here. If I want you, I'll call you.} This is akin to the two \textit{acceleration} and \textit{exploration} modes observed in prior work on AI-assisted programming~\cite{Barke2022GroundedCH}, indicating that planning suggestions might operate in a similar way as those of execution.

% \mgm{I like this section a lot! I'ts a bit sad it's last in teh who what where when why how ordering haha}

% \am{these two aren't about initiative per se their about expression modality}
Mediating these desires was the way in which the assistant was triggered.
For instance, analysts [A3, A8, A10] liked that the assistant responded to requests made in comments. 
% During the study, analysts naturally wrote comments down in their notebooks, which the assistant occasionally used as the context for suggestions. 
% Realizing this, some analysts continued requesting help in this way. A particularly illustrative example is when A8 typed \shortquote{assistant: how do I control for leagueCountry?} into a cell. Though our design primarily assumed assistant-initiated planning assistance, since some analysts showed a preference for their own initiative during the study, we treated analysts' comments as an initiative to elicit assistance.
% % enabled this behavior. \tim{what do we mean by enable?}
Some analysts wanted additional and different ways to directly ask the assistant for suggestions. A5, A6, and A7 wanted to ask the assistant for suggestions via a text prompt \`a la ChatGPT-style assistants.
% For example, A5 wondered if they \shortquote{could have typed something that says `plot discrete values scatter plot.'}  
Similarly, A11 wondered if the assistant could provide a list of relevant next steps via a keyboard shortcut.
% Designing an assistant to have a \am{various set} that can be adapted

This diversity of preferences suggests that having multiple ways to interact with the assistant that can be modified to taste and task is essential for effective assistant design. 
% , reminiscent of the functionality in Visual Studio Code \tim{did they mention this software specifically or is this functionality specific to that software (sounds surprising but I don't know)} where typing a keyboard shortcut on a specific part of the syntax yielded code auto-complete suggestions. 

%%%% 7-results.tex ends here %%%%

%%%% 8-discussion-future-work.tex starts here %%%%
\begin{table*}[t]
% \small
  \centering
  \renewcommand{\arraystretch}{1.23} % Adjust the vertical spacing of rows
  % \large
  \begin{tabular}{| p{2cm} | p{11.5cm} p{1.4cm}|}
    \hline
    & \textbf{Assistants should ...} & \\
    \hline
    \multirow{3}{*}{\parbox{2.0cm}{\textbf{Alignment}}} & 1. Communicate with analysts to align on analysis goals.&  (Sec.~\ref{sec:discussion_design_implications_match_goals}) \\
    % \cline{2-2}
    & 2. Provide suggestion content suited to the analysts' background.& (Sec.~\ref{sec:discussion_design_implications_match_bg}) \\
    % \cline{2-2}
    & 3. Match analysts' preferred level of planning assistance abstraction.& (Sec.~\ref{sec:discussion_design_implications_match_abstraction})\\
    \hline
    \multirow{2}{*}{\parbox{1.8cm}{\textbf{Timing and Initiative}}} & 4. Time suggestions based on analysts' openness to considering alternative approaches.& (Sec.~\ref{sec:discussion_design_implications_timing})\\
     & 5. Provide multiple ways to request content for different levels of abstraction. & (Sec.~\ref{sec:discussion_design_implications_initiative}) \\
    \hline
    \multirow{2}{*}{\parbox{2.0cm}{\textbf{Engagement and Trust}}} & 6. Mitigate suggestion overreliance by promoting engagement and critical thinking. & (Sec.~\ref{sec:discussion_design_implications_trust})\\
     & 7. Adopt multiple explanation methods for increased understanding and trust.
     % Mitigate suggestion overreliance by promoting engagement and critical thinking.
     % Encourage critical thinking and engagement to counterbalance suggestion automation.
     & (Sec.~\ref{sec:discussion_design_implications_explanation})\\
    \hline
  \end{tabular}
  \caption{Design implications and takeaways for building analysis assistants. }
  \label{tab:design_implications}
\end{table*}

\section{Discussion}
\label{sec:discussion}

In this work, we explored the potential of AI-based data analysis assistants that incorporate execution \textit{and} planning assistance.
 To identify the scope of suggestion content, we initially performed a \litreview review and a behavioral-driven analysis of independent analyses from a crowd-sourced analysis study (Sec. \ref{sec:suggestions}) to define categories of relevant suggestions (Table \ref{tab:suggestions}).
% To understand the space of helpful suggestion content, we perform a \litreview review and a behavior-driven study to establish a categorization of suggestion content. 
To explore the essential components of a helpful suggestion,  we conducted a \woz study to elicit the circumstances in which suggestions are helpful. 
% To explore the essential components of a helpful suggestion, we design and conduct a \woz study that is inspired by but not limited by the capabilities of existing LLM-based assistants. 
% This work contributes the first user study involving analysts working on a real dataset and task to better understand how to design and incorporate planning assistance into analysts' workflows. 
% In addition, we provide an open-source implementation of an assistant interface building on prior work. 
% We found that analysts saw many categories of planning assistance helpful, highlighting the potential of incorporating planning assistance. However, to achieve this, it is crucial to comprehend and address the contextual factors that influence analysts' willingness to consider planning suggestions. 

% Among these are aligning the analysis and the assistant on shared analysis goals and adding measures to reduce overreliance and promote trust in this new analyst-assistant paradigm. 
% \tim{what could we say here to be  more specific about engagement? promote trust but not overreliance? etc?}

We found that analysts had varying perspectives on what planning assistance was helpful. In some moments, planning assistance was about assistance for specific analysis steps; during others, it was about guidance for their overall workflow~(Sec. \ref{sec:results_what_assistance}). Some analysts found suggestions occasionally distracting, preferring to focus on their own analysis plan~(Sec. \ref{sec:results_when}). 
Similarly, we observed that there exists an inherent tension between the intended goals of the assistant and the goals of the analyst who might predominantly favor execution assistance, leading to divergent views on what constitutes ``helpfulness.''  Though analysts' goals are primary, assistants can play a crucial role in creating higher quality and more robust data analyses. Thus, supporting analysts with only execution assistance may not lead to such outcomes. Balancing analysts' perspectives of helpfulness and the assistant's impact on the workflow is crucial for providing truly valuable suggestions. Given that planning suggestions often require deeper consideration, a mutual understanding of their appropriate timing is also imperative. Developing affordances to enable analysts understanding and usage of such assistants is essential to those the design of those assistant. 
% An ``ideal'' assistant would completely understand the analyst's background, overall goals, and current plan (Fig.~\ref{fig:results}G and H) and reconcile this with its own goals (Fig.~\ref{fig:results}D). Further, the analyst should  understand the assistant's raison d'etre in addition to its capabilities. 
% Similarly, we found that while execution assistance offers time-saving advantages, it does not necessarily improve the robustness of analysis conclusions. Without well-reasoned planning considerations, execution assistance may focus on misguided directions resulting in wasteful efforts or misleading outcomes. 

% Beyond recommending analysts consider alternative decisions, data analysis assistants could prioritize helping analysts flesh out their analysis plan. Alternatively, assistants could only provide execution assistance in speeding up analysts' analysis plans. Or, the assistant could serve in a teaching role. 

Here, we synthesize these and other findings as design guidelines (Table~\ref{tab:design_implications}) to highlight design implications for future AI-based analysis assistants and reflect on limitations of our design.

\subsection{Design Guidelines}
\label{sec:discussion_design_guidelines}
Based on our results and discussion, we highlight seven design guidelines that extend Human-AI Interaction guidelines \cite{Amershi2019GuidelinesFH} for designing a data analysis assistant.

\subsubsection{Assistants should communicate with the analyst to align on analysis goals}
% The old  Creating common ground between analyst and agent could probably fit in here.
\label{sec:discussion_design_implications_match_goals}

% In this paragraph talk about how it should be on the agent to establish it? Or maybe in the parapgrah on how to establish

During the analysis portion of our study, the wizard was occasionally misaligned with the analyst's goals (Sec.~\ref{sec:results_when}). Likewise, analysts spent time figuring out the assistant's goals (Sec.~\ref{sec:results_who_curiosity}).
% \tim{it stands out to me that this is the "opposite direction"} 
% A few curious analysts typed comments seeking suggestions from the assistant and were curious about the assistant's resulting suggestions to understand the assistant (Sec.~\ref{sec:results_who_curiosity}).
% For example, participants noted if the assistant gave information similar to queries they had typed in comment boxes [A3, A8, A10]~\mgm{include A-X descriptions here if it's included in results section?}, and noted when the assistant did not respond to other kinds of artifacts in the notebook, such as error messages [A5, A13]. \ken{i usually just refer to results section but we can see what others think. Also I think this discussion was more about capabilities then goals.}
% Moreover, analysts in our study wanted more agency (e.g. directly prompting, choosing from a list of options, and specifying when they wanted to receive inputs) to direct the assistant toward their goals (Sec.~\ref{sec:results_how}).
Therefore, the analyst and assistant should establish shared goals at the beginning of the analysis, and this understanding should evolve over time. During its introduction, the assistant can describe its objectives for improving the analysis quality, highlight the value of planning assistance, and warn against the pitfalls of prioritizing execution assistance only. Furthermore, the assistant can communicate this information through interactive model cards~\cite{Crisan2022InteractiveMC}. Although prior work on conversational agents has found users prioritize their goals and needs over the assistant~\cite{clark2019makes}, we posit that it remains vital to consider both parties in the interest of quality analytical results. 
% In the context of data analysis, 
Adhering only to analysts' execution preferences may perpetuate poor analysis decisions and practices, leading to misguided inferences or biased results~\cite{mcnutt2020surfacing}. 
% ~(Sec.~\ref{sec:discussion_balance})
% Thus, it is necessary for an assistant to communicate clearly the goals of planning assistance to the analyst. Then, the analyst and assistant can coordinate the optimal collaboration. 
Understanding the best ways to establish, integrate, and uphold shared goals over time is an important area for future exploration.

% \tim{this is great, just make this a bit more clear above or argue for why it should only be in one section?}

\subsubsection{Assistants should provide suggestion content suited to the analyst's background}
\label{sec:discussion_design_implications_match_bg}

% different people had different prior knowledge, preffered one staistical approach over the other, had different levels of understanding, coding expertise etc. 
Analysts differ in their statistical, domain, and coding backgrounds. 
We observed that the same suggestion considered helpful by some participants, could be found by others to be too elementary, too distracting, too far from their expertise (making correctness auditing difficult), or too different from their usual analysis or coding practices to be useful~(Sec.~\ref{sec:results_who}). 
% We observed that these differences affected their analysis plan and reactions to suggestions~(Sec.~\ref{sec:results_who}). The same suggestion considered helpful by some participants, could be found by others to be too elementary, too distracting, too distinct from their expertise (making testing its validity was prohibitively difficult), or too different from their usual analysis or coding practices to be useful (Sec. \ref{sec:results_who_expertise}). 
% Thus, assistants should match suggestion content to the analyst's background.
In situations when a suggestion is too difficult or different to comprehend, the assistant should provide options to explain further or offer additional resources---e.g., links to relevant documentation and tutorials. Likewise, the assistant could suggest  more approachable alternatives on demand. 
% These additional explanations or approaches could be visible through a show/hide toggle.
% \tim{just a comment: I like allowing the user to expand/contract details rather than trying  to have the assistant always get it right. just allow them to hide or ask for more explanation. }
 % Prior work on conversational agents found many users thought the system should take on the burden of establishing common ground to personalize outputs~\cite{clark2019makes}\mgm{This seems to be an interesting point but I'm not quite sure where to put it.}. 
In our study, the assistant did not gather any information on task or background from the analyst. A more explicit definition of the analyst's expertise, such as through analyst initiative (Fig. \ref{fig:results}A), may help establish common ground.
For instance, on first-usage analysts could have a quick conversation with chatbot to elicit their educational background, coding competency, and typical workflows (akin to how we asked participants about their preferred environments before the study). 
% For instance, analysts could configure their backgrounds by providing a short bio 
% \tim{do we really think a bio would be the right thing to use?  more specific questions seem more targeted and more likely to succeed?} 
% or a completed template describing their educational background, a self-rating of their coding expertise, typical workflows, and preferences for analysis methods and software packages. 
The assistant could then use this as relevant context to customize LLM-based assistant suggestions \cite{Andreas2022LanguageMA, Kirk2023PersonalisationWB}. 

% different levels of explanation for different audiences
\subsubsection{Assistants should match the analyst's preferred level of planning assistance abstraction.}
\label{sec:discussion_design_implications_match_abstraction}

There are multiple perspectives on what consitutes helpful planning assistance. These can range from guiding hyper-parameter selections to providing a comprehensive plan with multiple analysis steps.
Future assistants should account for analysts' preferences regarding desired levels of planning assistance and how these preferences may change throughout the analysis process.
% The assistant could learn analysts' preferences by observing their behavior across multiple analyses and within single analysis, goals may different between sessions. 
% More simply, this could be communicated explicitly by the analyst in the UI. 
% For instance, analysts may want to iteratively refine their intent for assistance such as in \textit{Wrex}, a non-LLM-based notebook extension that generates code through programming by example \cite{Drosos2020WrexAU}.
For instance, the assistant could offer explicit modal dials to guide suggestion content---such as an ``\textit{execution}'' mode that focuses on code execution, a ``\textit{think}'' mode for specific planning suggestions, a ``\textit{reflection}'' mode for connecting decisions and highlighting potential missed steps, and an ``\textit{exploration}'' mode for higher-level planning suggestions.
% ---although such a partitioning would need to be designed to be in accord with the interleaved and iterated structure of data analysis.
Likewise, the level of planning abstraction also relates closely to the scope of the analysis code that the assistance affects. An analytical decision could be one parameter, line, function, code block, or even multiple chunks of code \cite{Liu2020BobaAA}. Suggestions that impact large parts of the analyst's analysis require more willingness from the analyst to consider and adopt. 
Both sets of levels could be configured from the UI or learned through observation of the users across analyses.

\subsubsection{Assistants should time suggestions based on the analyst's openness to considering alternative approaches.}
\label{sec:discussion_design_implications_timing}

Analysts  in our study felt that some suggestions were good but were not well-timed based on their current task~(Sec. \ref{sec:results_when}). 
When analysts are focused on executing a specific analysis decision or plan, they may prefer execution rather than planning suggestions. Conversely, when they are considering their next steps, they may be more open to guidance for analysis planning.
% Therefore, assistants should strive to recognize when analysts are open to planning assistance. 
Analysts could explicitly communicate their timing preferences to the assistant or use a ``remind me later'' response to raised suggestions. Additionally, organizing and tagging suggestions could help them easily find and revisit suggestions.
% even if their timing were not ideal. 
 Meanwhile, the assistant could also learn appropriate timing from analysts' behaviors. For instance, prior works have used probalistic utility modeling~\cite{Horvitz1999PrinciplesOM} and sensors to model and predict human interruptibility~\cite{Fogarty2005PredictingHI}. 
 % The assistant could also follow a utility model for mixed-initiative AI intervention by considering the expected utility of different actions~\cite{Horvitz1999PrinciplesOM}. 
We note that this may be challenging if analysts fixate on rigid analysis plans and highlight the importance of establishing shared goals between the analyst and assistant (Sec. \ref{sec:discussion_design_implications_match_goals}). 
% While many analysts may prefer planning assistance, others may not. Given this tension and difficulty, considerations for the timing of suggestions may also benefit from the recommendations of \textit{polite computing}~\cite{Whitworth2005PoliteC}.
% \tim{the last sentence speaks to the question: Is there ever a good time to engage the analyst on planning assistance. It sounds  like while this is the case for many, it isn't for some and that's an inherent tension. perhaps clarify that slightly here.}

\subsubsection{Assistants should provide multiple ways to request content at different levels of abstraction.}
\label{sec:discussion_design_implications_initiative}

We observed diverse opinions (Sec.~\ref{sec:results_what_assistance}) about assistant-versus-analyst initiative and levels of planning assistance, a heterogeneity which might be met by supporting multiple means of requesting assistance.
% Though preferring assistant-initiated suggestions, analysts also wanted to exert their own initiative in requesting suggestions (Sec.~\ref{sec:results_how}). 
% \jeff{I'm a bit thrown off by this, as I thought we said earlier that participants liked having suggestions initiated by the system} 
% Further, they showed different preferences for the level of planning assistance (Sec.~\ref{sec:results_what_assistance}). Desires which would be supported by support multiple means of requesting assistance.
% Therefore, assistants should provide analysts with multiple ways to request assistance for different levels of abstraction. 
% In terms of design, there is flexibility in the input mechanism and initiative. For example, analysts could write comments, select text, trigger suggestions via a hotkey, request proactive suggestions in a side panel, etc. Each could carry implicit context regarding the level of feedback they desire \cite{McNutt2023OnTD}. Ambient and textual interfaces (i.e., assistant follows the cursor and can be invoked via an action) suggest localized context and help. For example, a common workflow with execution assistants such as Github Copilot has the programmer write comments that detail what the lines of code should be to complete the task \cite{Vaithilingam2022ExpectationVE}. Meanwhile, a side-panel interface with controls outside the immediate notebook suggests global context and assistance. 
While existing systems usually offer one form of invocation, LLM-based analysis assistants should allow and interleave multiple forms of invocation and context for different levels of assistance. It is also essential to clarify the context for the LLM in each form of interaction. For example, a hotkey trigger at the line level could indicate a focus on execution assistance using only the current cell. On the other hand, analysts could specify broader suggestions when requesting feedback in the side panel and choose the LLM context by selecting relevant cells. 
Interactive visualizations could also be employed to show and help analysts make decisions and consider alternative choices \cite{Liu2019PathsEP, Liu2020BobaAA}. These visualizations (e.g., Fig~\ref{fig:teaser}), tied to the analysis code, could help analysts review their steps and select areas where additional assistance is needed.

\subsubsection{Assistants should mitigate suggestion overreliance by promoting engagement and critical thinking.}
\label{sec:discussion_design_implications_trust}

% Results relations
    % Errors exist -- and this is a known problem
% LLMs are adaptable and powerful, but they can have erroneous outputs~\cite{Amershi2019GuidelinesFH,bender2021dangers}. 
% As data analysis often informs high-stakes decisions~\cite{Baker20161500SL, Cockburn2020ThreatsOA, Aarts2015EstimatingTR}, it is critical for analysts to be aware of the \am{wc}decisions and rationales that comprise their analysis even as the assistant makes seemingly sensible recommendations---particularly given LLM's hallucinatory tendencies~\cite{Amershi2019GuidelinesFH,bender2021dangers}. 
As data analysis often informs high-stakes real-world decisions~\cite{Baker20161500SL, Cockburn2020ThreatsOA, Aarts2015EstimatingTR}, it is essential that analysts be able to understand and think critically about their analyses, regardless whether they received execution or planning assistance.
% Likewise, different individuals' trust in automation can impact their response to erroneous outputs~\cite{lee2004trust,zhang2020effect}. 
% skepticism towards the suggestion quality and 
In our study, analysts showed varying levels of  thoroughness in validating the model's decisions, code, and outputs (Sec.~\ref{sec:results_who_curiosity})---sometimes critically disengaging and going on \shortquote{autopilot}. 
% % For example, A4 caught an error in which a ChatGPT-generated suggestion described a p-value of $0.15$ as significant. 
%     % Varying trust levels and understanding of LLMs
% For instance, A8 \shortquote{acted on autopilot} over-relying on the assistant's suggestions without critical engagement (Sec.~\ref{sec:results_why}), while A4 caught an incorrect LLM-based suggestion that p-value of 0.15 is significant.
% Gordon et al. \cite{gordon2023co} sketch the growing area of tools for auditing AI generate code output. 
% Therefore, assistants should reduce overreliance by introducing mechanisms to promote engagement and critical thinking with their suggestions.
% One way to increase engagement is to provide explanations that make it easier to understand and verify the assistant's suggestions~\cite{vasconcelos2023explanations}. 
 Over-reliance on AI is well documented in other scenarios ~\cite{buccinca2021trust,bussone2015role,vasconcelos2023explanations,yin2019understanding}, and developing mechanisms for skepticism is a growing concern~\cite{gordon2023co}.
% We might 
Designing assistants with critical affordances is essential to preventing them becoming AI-mediated p-hacking machines. 
% Embracing some of these approaches is of critical importance \am{to making our assitant a good boy}
One approach might be to reduce the cognitive effort of engaging with suggestions~\cite{vasconcelos2023explanations}, freeing analysts to be more scrutinous.
% may be less prone to accept the assistant's recommendation without scrutiny~\cite{vasconcelos2023explanations}. 
% \tim{I wonder whether that's always true. You could reduce cognitive effect with an 1click "accept and implement" but this could of course backfire. is it perhaps not only about cost/effort? your next paragraph suggests that as well}\ken{just talking about explanations and not about adopting the suggestions -- changed from "are" to "may be"} 
For example, the system could present a visualization mapping the overall decision paths (\`{a} la Fig.~\ref{fig:teaser})  to demonstrate how a given recommendation would influence the overall analysis.
% The visualization could highlight decisions that would be affected or illustrate alterations in the decision paths. 
% Similarly, the assistant can reduce the cost of contemplating their decisions and invite critical thinking by providing a series of reflective questions~\cite{Wood2019DesignEW}. \tim{is that really reducing cost? or increasing it if we ask more? it might depend on what we view as the alternative/counterfactual. it's more work compared to not doing anything or simply accepting it. but it's less cost relative to having to think through all of this carefully without helpful scaffolding. I think readers could get confused here.} 
% Additionally, incorporating multiple explanation methods could improve comprehension among analysts, a design guideline we further elaborate on in Sec.~\ref{sec:discussion_design_implications_explanation}.
% Besides simplifying the process of understanding the assistant's suggestions, 
Similarly, assistants could introduce cognitive forcing functions~\cite{buccinca2021trust} to increase the cost of relying on the assistant. For example, the assistant could require a  scaffold of the analysis plan before providing suggestions or by requiring that code be typed out manually rather than automatically inserted. 
Forced low-level participation may prompt better engagement with the assistant---although it is crucial to ensure that these interventions do not deter analysts from utilizing assistance altogether.

\subsubsection{Assistants should adopt multiple explanation methods for increased understanding and trust.}
\label{sec:discussion_design_implications_explanation}
Analysts highly valued suggestion explanations, as they helped them better understand and trust the assistant's suggestions (Sec.~\ref{sec:results_what_qualities}). 
% In our study, code, code comments, and natural language suggestion rationales all contributed to explaining suggestions. 
Explanations provided both rationale for why an analysis decision should be considered and definitions of potentially unfamiliar concepts. 
Future assistants should incorporate multiple forms of explanations, that are sensitive to  analysts' backgrounds and preferences. To avoid introducing excessive cognitive load~\cite{vasconcelos2023explanations}, assistants could allow analysts to select their preferred types of explanation. 
Alternative explanation mediums, like visualizations or animations~\cite{Pu2021DatamationsAE, Xiong2022VisualizingTS}, could better contextualize suggestions.
% other potential methods to present suggestion explanations. For example, visualizations or animations of the data during and after data transformations could help analysts understand the assistant's code actions~\cite{Pu2021DatamationsAE, Xiong2022VisualizingTS}. 
% Recent work~\cite{Dibia2023LIDAAT, Lin2022TeachingMT, Liu2023GEvalNE} has demonstrated that LLMs, like GPT-4, can assess the quality of their output, approaching human-level judgment for tasks such as summarization and visualization.
% which might be useful for constructining explanations.
% \tim{that's a pretty strong claim? I didn't look at these papers but I imagine they couldn't always do that. otherwise they wouldn't need people anymore :) I think we should probably make this softer of more nuanced.}
As LLMs continue to improve, assistants could also provide model-generated critiques of their own suggestions~\cite{Dibia2023LIDAAT, Lin2022TeachingMT, Liu2023GEvalNE}.
% as an initial pass at explaining the strengths and limitations of an approach. 
% Nevertheless, 
Identifying explanation best practices remains an open area of research~\cite{bansal2021does,buccinca2021trust,eiband2019impact,vasconcelos2023explanations}. 
% Future work could explore the effects of different explanation types in the context of a data analysis task. 
Unlike many other tasks however, there are often multiple reasonable paths in data analysis~\cite{Liu2019PathsEP} without an objectively ``correct'' answer or other measures besides accuracy---making evaluation of explanations especially troublesome.
% ---such as alternative paths considered, analysis quality (though this is difficult to evaluate), or analysts' perceptions and actions---could help assess different explanation designs. 
% Although this lack of an objective ``correct'' answer deviates from prior work, it may also open to the door to interesting new analyses.  
% Furthermore, future studies can still measure the impact of explanation design on analyst perceptions and actions. 
% For example, a study could present incorrect suggestions and measure if analysts notice the errors under different explanation conditions.
% Future work could also apply similar methods to judge the effectiveness of the assistant intervention through the quality of the final analysis outputs, or by creating experimental conditions in which the incoming dataset has flaws and seeing if analysts perform measures to address those flaws.

% \tim{I don't like when limitations is the very last part of the paper. so much good work to end on a concentrated note of all the potential issues. I think I'd either prefer a very short conclusion back or having some kind of other last subsection here.}

\subsection{Limitations and Future Work}
\label{sec:discussion_limitations}
This work has several limitations, relating to our participants, our study design, and our analysis of the results. 

% \am{unhappy with this paragraph}
First, we selected analysts who self-identified as having a high proficiency in statistical analysis. As we wanted to understand the mechanisms surrounding planning assistance, we determined that analysts with sufficient programming and statistical knowledge would be more engaged with analysis planning and our assistant's suggestions; we found that even they were often unaware of the decisions and rationales our assistant offered. 
% As such analysts are often the ones performing analyses that influence real-world decisions, understanding planning assistance with this audience is paramount. 
Analysts with less statistical and programming expertise may have different experiences and preferences when working with an analysis assistant. Future work should explore how this style of assistance can be support for novice data analysts.
In addition, we focused on people willing to participate and perform data analysis with an AI assistant, which may positively bias our results.
% This may affect analysts' willingness to work with an assistant and eagerness to probe the assistant's capabilities. 

Next, to facilitate a lab study of reasonable duration, we chose to conduct a same-day, in-person study of two hours. 
% If analysts had more time with the assistant, say weeks, they may have calibrated their expectations for the behaviors of the assistant and developed a better sense of when to rely or not to rely on it. Likewise, analysts may have developed a better sense of the assistant's timing and goals, as well as overcome any potential novelty effects latent to our one-shot design.
If analysts had more time with the assistant, say weeks, they may have calibrated their expectations for the behaviors of the assistant, developed a better sense of the assistant's timing and goals, and overcome any potential novelty effects latent to our one-shot design.
In future work we intend to explore these issues by developing a tool that supports this style of longitudinal usage.
% , which can be realistic in certain scenarios,
Similarly, while our lab-based \woz study design afforded us substantial control over the assistant, it may not have presented the same stakes as a real-world one. 
% Outside the lab, conclusions from analyses may impact critical decisions, and analysts may be more careful because of this. 
% Similarly, the time pressure may have pushed analysts towards favoring execution assistance. 
As a result, participants may have been more willing to accept planning and execution assistance. We tried to mitigate this effect by priming analysts to consider the robustness of their analyses via our introductory material. 
% , and, contrastingly, may have placed assistant suggestions under greater scrutiny

% —when provided with planning assistance—
Finally, our study concentrated on analyst preferences, leaving evaluation of the resulting quality of analyses unaddressed. Determining analysis correctness is inherently challenging \cite{Silberzahn2018ManyAO, Breznau2022ObservingMR, Schweinsberg2021SameDD, Dutilh2018TheQO, BotvinikNezer2019VariabilityIT, Bastiaansen2019TimeTG, Menkveld2021NonStandardE}, with there often being 
% given that real-world analytical questions can yield a range of defensible analysis approaches \cite{Silberzahn2018ManyAO, Breznau2022ObservingMR, Schweinsberg2021SameDD, Dutilh2018TheQO, BotvinikNezer2019VariabilityIT, Bastiaansen2019TimeTG, Menkveld2021NonStandardE}. 
% Often, there is 
no single ``correct'' answer, but rather multiple defensible interpretations. However, extensive literature on multiverse analysis \cite{Steegen2016IncreasingTT, Simonsohn2015SpecificationCD, Liu2020BobaAA} notes the importance of considering alternative approaches. This helps analysts appreciate uncertainties surrounding their decisions and enhances analysis robustness. 
This work takes a step toward this larger goal by showing how planning assistance can guide analysts toward assessing multiple analytical perspectives. 
\begin{acks}
We are grateful for the participants of our study and thank the UW Behavioral Data Science Group members for their suggestions and feedback. We also thank Tiffany Zheng for her brainstorming on the figures and moral support. T.A. and K.G. were supported in part by NSF grant IIS-1901386, NSF CAREER IIS-2142794, and the Bill \& Melinda Gates Foundation (INV-004841).
\end{acks}

%%
%% The next two lines define the bibliography style to be used, and
%% the bibliography file.

\bibliographystyle{ACM-Reference-Format}

%%%% bibliography starts here %%%%

%%% -*-BibTeX-*-
%%% Do NOT edit. File created by BibTeX with style
%%% ACM-Reference-Format-Journals [18-Jan-2012].

%\bibliography{0-references_custom}%% Commented by merge tool

% \printbibliography

%%
%% If your work has an appendix, this is the place to put it.

\input

%%%% 10-appendix.tex starts here %%%%

\clearpage
\appendix

\section{Examples of Suggestion Categories}
\label{appendix:suggestion_examples}

We give a brief description of each category in Table~\ref{tab:suggestions} and illustrate full examples of suggestions for each of the suggestion categories. 

\subsection{Domain Background}
 While domain background content could be part of suggestions in other categories, this category is for when the main focus of the suggestion is to provide analysts with background on the problem domain. 
 An example could be giving analysts suggestions about similar analyses done in the past:
\begin{itemize}
  \item \xhdr{Similar Analyses Done in the Past} 
  
  The question of whether skin tone influences the way referees make decisions in soccer games has been a topic of research for a while. One of the most well-known studies in this area is "Racial Bias in the Allocation of Fouls in Soccer" by Price and Wolfers (2010). They analyzed data from the 2004-2005 English Premier League season and found that black players were more likely to receive red and yellow cards than white players, even after controlling for a variety of factors such as position, nationality, and the competitiveness of the game.

    Another study that investigated the issue is "Skin color and red card decisions in football: a preliminary study" by Diaz et al. (2016). They analyzed data from the Spanish league and found that darker-skinned players were more likely to receive red cards than lighter-skinned players, after controlling for various factors such as age, nationality, position, and the competitiveness of the game.

    Both studies used regression analysis to examine the relationship between race and the likelihood of receiving red cards. Price and Wolfers included a wide range of control variables, while Diaz et al. included fewer variables but also examined the moderating effect of the referee's country of origin.

    Overall, the analyses focused on controlling for various factors that could influence the relationship between race and red cards, and both studies found evidence of racial bias in the allocation of fouls.
\end{itemize}

\subsection{Data Wrangling}
Offer assistance related to understanding, transforming, clearing, and preparing the data.

Example suggestions: 
\begin{itemize}
  \item \xhdr{Reminder - Check for Missing Data}
  
  Always check for missing data in your DataFrame before performing any data analysis or machine learning tasks. NaN values can cause unexpected errors and affect the accuracy of your results.
    \begin{lstlisting}[caption={}, label={lst:pythoncode}]
# Use boolean indexing to filter the DataFrame and show the rows with missing data
rows_with_na = df[df.isna().any(axis=1)]
rows_with_na.head()
# Show how many NaN values there are in each column
nan_counts = df.isna().sum(axis=0)
    \end{lstlisting}
  \item \xhdr{Understanding Differences Between Raters} 
    
    To help us understand the differences between raters better we can use the following code:
    \begin{lstlisting}[caption={}, label={lst:pythoncode}]
# Compute the difference between the two raters' ratings}
data['rating_diff'] = data['rater1'] - data['rater2']
# Create a distribution plot of the rating differences using seaborn
sns.histplot(data=data, x='rating_diff', kde=True)
    \end{lstlisting}

    \item \xhdr{Data Transformation - ZTransform Predictors}

    In this dataset, we may want to z-transform predictors as it can improve the interpretability of regression coefficients by putting all predictors on the same scale. Standardization can also reduce the impact of outliers and extreme values that can skew the results of statistical analyses. Finally, standardization can make it easier to compare the relative importance of predictors in a model that has different units and scales.

    A code template to z-transform a variable is shown:
    \begin{lstlisting}[style={rstyle}, caption={}, label={lst:rcode}]
df$z_var <- scale(x=df$var, center=TRUE, scale=TRUE)
    \end{lstlisting}
\end{itemize}

\subsection{Conceptual Model Formulation}
Inspired by Jun et al.~\cite{Jun2021HypothesisFE}, this category corresponds to assistance that helps analysts identify or think about factors that might influence an outcome and how these factors might interact to give rise to the outcome. 

Example suggestions:
\begin{itemize}
  \item \xhdr{Age Variable Rationale}
  
  Beyond just skin tone we might want to consider other variables that might affect the analysis. As birthday is in the dataframe, we might want to control for age which is a common practice in research studies, as age can affect various aspects of behavior and cognition. One reason to control for age is to account for the tendency of impulsivity, which may be associated with receiving red cards, to decrease with age. Research has shown that sensation seeking and impulsivity tend to decrease with age, as indicated by both self-report and behavior measures. Therefore, controlling for age can help reduce this variable's impact on the study results. Additionally, age is a common demographic variable, and controlling for it can help to ensure that any observed effects are not simply due to age differences among participants.

    We can create a new column \emph{age} based on the \emph{birthday} column, we first need to convert the \emph{birthday}  column to a date format using the \emph{as.Date} function. We can then calculate the age using the \emph{difftime} function, which calculates the time difference between two dates in years.
    \smallskip
    \begin{lstlisting}[style={rstyle}, label={lst:pythoncode}]
# convert birthday column to date format
df$birthday <- as.Date(df$birthday, format = "%d.%m.%Y")
# calculate age as of Jan 1, 2013
df$age <- floor(as.numeric(
        difftime(as.Date("2013-01-01"), df$birthday, units = "days")
    ) / 365.25)
    \end{lstlisting}
  \item \xhdr{ New Variable Star Status and Rationale} 
    
    We can create a variable to capture star status. More popular or famous players might potentially be given more lenient treatment by referees and be less likely to be ejected with a red card. Based on the data available, we can determine stat status by calculating the average number of goals scored by each player per game they played in the data set.

    Although this is imperfect in that it might not properly account for players who are defensive stars and score fewer goals, offensive players are typically the most famous, and thus goals scored should generally be a meaningful indicator of this status.

    The code below shows how we might do this:
    \begin{lstlisting}[style={rstyle}, label={lst:pythoncode}]
# Group by playerShort and sum goals and games
grp <- aggregate(df[c("goals", "games")],
by=list(df$playerShort), sum) %>%
rename(playerShort = Group.1)
# Calculate goals per game
grp$goals_per_game <- grp$goals / grp$games
# Merge the goals per game back into the original dataframe
df <- merge(df, grp[c("playerShort", "goals_per_game")], by="playerShort")
    \end{lstlisting}

    \item \xhdr{Other Variables that Might Influence Red Cards: Physical Stature and Rationale}

As we are looking at height and weight, we may want to consider physical stature as one variable. More physically imposing players could be more likely to be perceived as committing a violent act and thus earn more red cards. 
Because all players are elite professional athletes, we can use BMI as it is likely to reflect athletes having a larger and more muscular frame rather than being more overweight as in the general population.

To create a new column bmi based on the height and weight columns, we can use the formula for calculating BMI, which is the weight (kg) divided by height (m) squared. 
    \begin{lstlisting}[caption={}, label={lst:rcode}]
# convert height from cm to m
df['height_m'] = df['height'] / 100
# calculate bmi
df['bmi'] = df['weight'] / (df['height_m'] ** 2)\end{lstlisting}
\end{itemize}

\subsection{Operationalizing Constructs}
Help analysts with operationalizing variables involved in the analysis. This may involve data transformations but if the primary goal is to create new columns that can be worked with in the analysis, then the suggestion is about \textit{operationalizing constructs}.

Example suggestions: 
\begin{itemize}
  \item \xhdr{Skin Tone Operationalization}
  
  As you are exploring skin tone ratings, several other methods could be used. The code below shows a basic template to do the above ways of representing skin tone.
    \begin{lstlisting}[caption={}, label={lst:pythoncode}]
# create a new column for average rating
df['avg_rating'] = (df['rater1'] + df['rater2']) / 2
# create a new column for minimum rating
df['min_rating'] = df[['rater1', 'rater2']].min(axis=1)
# create a new column for maximum rating
df['max_rating'] = df[['rater1', 'rater2']].max(axis=1)
# create a new column for dark skin tone as a cutoff
df['dark_skin_tone'] = (df['avg_rating'] > 0.75).astype(int)
# view the first 10 rows of the new columns
df[['avg_rating', 'min_rating', 'max_rating', 'dark_skin_tone']].head(10)
    \end{lstlisting}
  \item \xhdr{Collapsing the Data by Player} 
    
    Based on the dataset description, each player-referee dyad has multiple games in which the player and referee appeared. As we explore rating distributions, it may be reasonable to collapse the data across \emph{playerShort}, keeping the ratings for the player. We can also create a variable \emph{sumGames} that are aggregated across the variable games that provide the summed number of games played by the player across all player-referee dyads. For the outcome variable, we can create \emph{sumRedCards} that represents the number of red cards a player received from all referees encountered.

    The code below does this transformation:
    \begin{lstlisting}[caption={}, label={lst:pythoncode}]
# Define the columns to take the first value of
columns_first = ['rater1', 'rater2', 'position', 'avg_rating', 'height', 'weight', 'birthday', 'leagueCountry', 'refCountry']
# Define the columns to sum and rename them
columns_sum = ['games', 'redCards', 'victories', 'goals']
columns_sum_renamed = [f'sum{x.capitalize()}' for x in columns_sum]
# Group the data by player and get the first value of specified columns
# and the sum of specified columns
first_cols = df.groupby('playerShort')
first_cols = first_cols[columns_first].first()
sum_cols = df.groupby('playerShort')
sum_cols = sum_cols[columns_sum].sum()
# Add a column 'dyads' with the number of dyads each player belongs to
sum_cols['dyads'] = df.groupby('playerShort').size()
# Concatenate the two data frames and reset the index
player_df = pd.concat([first_cols, sum_cols], axis=1).reset_index()
# Rename the columns
column_names = (['playerShort'] + columns_first + columns_sum_renamed + ['dyads'])
player_df.columns = column_names
player_df.head()
    \end{lstlisting}
\end{itemize}

\subsection{Choosing the Statistical Model}
Analysis assistance which aids analysts in choosing or directly points analysts to a relevant statistical model. 

Example suggestions:
\begin{itemize}
  \item \xhdr{Understanding Differences Between Raters} 
    
    We can start with a simple OLS model that includes the player's skin tone and the number of red cards as the dependent variable. 

    Here is an example of how we can create this model in R:
    \begin{lstlisting}[style={rstyle}, label={lst:pythoncode}]
# Create an OLS model
model <- lm(redCards ~ avg_rating, data = df)
# Summarize the model
summary(model)
    \end{lstlisting}
    
  \item \xhdr{Modeling Strategy: Generalized Linear Models}
  
Beyond just using linear models, we may want to account for the random variance of the effects across players, referees, and referees' country of origin. We could use generalized linear mixed models (function \textit{glmer} in R package \textit{lme4}).

 To start, you could fit three models of increasing complexity to explore the effects of skin tone on the likelihood of receiving a red card. An example template of this code is shown below:
 
 \begin{lstlisting}[style={rstyle}, label={lst:pythoncode}]
gm1 <- glmer(redCards ~ 1 + avg_rating + (1 | playerShort) + (1 | refNum), family = binomial, data = df)
gm2 <- glmer(redCards ~ 1 + avg_rating + (1 | playerShort) + (1 + avg_rating | refNum), family = binomial, data = df)
gm3 <- glmer(redCards ~ 1 + avg_rating + (1 | playerShort) + (1 | refNum) + (1 + avg_rating | refCountry), family = binomial, data = df, nAGQ = 0)
\end{lstlisting}

\emph{gm1} assumes that the effect skin tone has on the likelihood of receiving a red card is constant across all referees and players, but that the baseline rate of red cards varies across referees and players. \emph{gm2} allows the effect skin tone to vary across different referees, and \emph{gm3} further allows for systematic differences in the way that referees from different countries make biased decisions based on skin tone.

All models assume that observations are independent and identically distributed (i.i.d.), the response variable follows a binomial distribution, and the random effects follow a normal distribution. They also assume that the random intercepts and slopes are uncorrelated with each other and with the fixed effects.

Further adjustments can be made as appropriate, such as changing the family type or adding other fixed effects.

    \item \xhdr{Logistic Regression with Statsmodels}
    
As the red cards are converted to a binary variable we can run a logistic regression using the statsmodels library. Here's some sample code to run:
    \begin{lstlisting}[caption={}, label={lst:rcode}]
import statsmodels.api as sm
X = sm.add_constant(X)  # Add an intercept term to the model
model = sm.Logit(y, X).fit()
print(model.summary())
\end{lstlisting}
The resulting output will give you information about the coefficients for each variable in the model and whether they are statistically significant predictors of receiving a red card, while controlling for other variables in the model. You can also use the output to calculate odds ratios to quantify the association between skin tone and receiving red cards.

\end{itemize}

\subsection{Model Results Interpretation}
This category involves any suggestion that helps interpret the outputs of running a statistical model into a more readable and potentially actionable summary. These suggestions are primarily written by ChatGPT

Example suggestions:
\begin{itemize}
%   \item \xhdr{OLS Results Interpretation} 
    
% The R-squared value is 0.018, indicating that the independent variables explain only 1.8\% of the variation in the dependent variable. The adjusted R-squared value is 0.008, which takes into account the number of independent variables in the model and penalizes the R-squared value for using too many variables.

% The F-statistic tests whether there is a relationship between the independent variables and the dependent variable. The p-value for the F-statistic is 4.94e-06, indicating that there is a statistically significant relationship between the independent variables and the dependent variable.

% The coefficients for the individual clubs indicate the effect that each club has on the number of red cards. A positive coefficient means that players from that club receive more red cards, while a negative coefficient means that players from that club receive fewer red cards, compared to the average. However, many of the coefficients have p-values greater than 0.05, indicating that they are not statistically significant. Therefore, we cannot conclude that these clubs have a significant effect on the number of red cards received by players.
    
  \item \xhdr{ANOVA Results Interpretation}
  
The output shows the results of a one-way ANOVA test that examines the relationship between \emph{rater1} and \emph{redCards}. The \emph{redCards} variable is grouped by the levels of \emph{rater1}, and an ANOVA test is performed to determine whether there is a statistically significant difference in the mean \emph{redCards} scores between these groups.

The F-statistic of 2.35 indicates the ratio of the variance between the groups to the variance within the groups. The p-value of 0.0523 indicates the probability of observing such an F-statistic value or higher under the null hypothesis that there is no significant difference in the mean \emph{redCards} scores between the groups.

Based on the p-value of 0.0523, we can conclude that there is a borderline statistically significant difference in the mean \emph{redCards} scores between the groups defined by the \emph{rater1} variable. Specifically, we can reject the null hypothesis at a significance level of 0.05, but not at a significance level of 0.01. Therefore, there is some evidence to suggest that the \emph{rater1} variable is related to the \emph{redCards} variable, but the evidence is not very strong.
\bigskip
    \item \xhdr{Comparing and Interpreting OLS Results}
    
Looking at the three OLS regression results, we can see that the models have low R-squared values, indicating that the predictor variables explain very little of the variance in the dependent variable, \emph{redCards}. Additionally, the F-statistic for each model is relatively low, indicating that the models may not be a good fit for the data.

Model 1 has two predictor variables, \emph{skinTone} and \emph{goals}, with \emph{goals} having a higher coefficient than \emph{skinTone}.

Model 2 has three predictor variables, \emph{skinTone}, \emph{age}, and \emph{Intercept}, with \emph{Intercept} having the highest coefficient, followed by \emph{age} and \emph{skinTone}.

Model 3 only has two predictor variables, \emph{skinTone} and \emph{Intercept}, with \emph{Intercept} having a higher coefficient than \emph{skinTone}.

Given that the models have low R-squared values, it may be worth trying a different model or adding additional predictor variables to better explain the variance in the dependent variable. Additionally, it would be helpful to investigate the relationship between the predictor variables and the dependent variable further to better understand their impact on the model.

\end{itemize}

\subsection{High-Level Planning}
These are suggestions that point analysts to a general analysis decision, have a high level of abstraction, and are not specific or easily actionable enough to fit into the other planning categories discussed before. 

Example suggestions:
\begin{itemize}
  \item \xhdr{Consider Including Covariates}
  
Covariates are additional variables that can affect the relationship between the predictor variable and the outcome variable in an analysis. In order to accurately understand the relationship between the predictor and outcome variable, it is important to consider the effects of these covariates.

Including covariates in an analysis can help to control for potential confounding variables, which are variables that can affect both the predictor and outcome variables. By controlling for these confounding variables, the true relationship between the predictor and outcome variable can be more accurately estimated.

Additionally, including covariates can help to increase the precision of the estimated effect of the predictor variable on the outcome variable. This can lead to more accurate predictions and better decision-making based on the analysis results.

In summary, considering covariates in an analysis is important because it helps to control for potential confounding variables and increase the precision of the estimated effects of predictor variables on outcome variables.
  \item \xhdr{Visualization Suggestions} 

Based on the dataframe provided, here are some visualizations that can be considered to understand the relationship between variables:

1.  Histogram: A histogram of the \emph{redCards} variable can be created to see the distribution of the number of red cards given by referees. This can help identify if there are any extreme values or outliers in the dataset.
    
2.  Scatter plot: A scatter plot of \emph{redCards} against \\ \emph{Dark\_Skin\_Cards} can be created to visually see if there is any relationship between the two variables. This can help identify if there is any correlation between a player's skin tone and the number of red cards they receive.
    
3.  Box plot: A box plot of \emph{redCards} grouped by \\ \emph{Dark\_Skin\_Cards} can be created to see if there is any difference in the distribution of red cards received by players with different skin tones.
    
4.  Bar chart: A bar chart of the total number of red cards given by each referee can be created to see if there are any referees who give out a disproportionately large number of red cards.
\end{itemize}

\subsection{Execution Assistance}
This is a broad category to cover any suggestion that helps analysts carry out an analysis decision. This can involve providing code, aiding analysts in debugging errors, or helping implement analysts' intent. These suggestions are also primarily created using ChatGPT. 

Example suggestions:
\begin{itemize}
  \item \xhdr{Correlation Calculation}
 
You can calculate the correlation coefficient between the two variables to see the correlation between skin tone and red card. Here's how you can do it in Python using the \emph{pearsonr} function from the \emph{scipy.stats} module:
    \begin{lstlisting}[caption={}, label={lst:pythoncode}]
from scipy.stats import pearsonr
# Calculate Pearson correlation coefficients and p-values for rater
corr_rater, pval_rater = pearsonr(X['rater'], X['redCards'])
print(f"Correlation coefficient (rater): {corr_rater:.2f}")
print(f"p-value (rater): {pval_rater:.2f}")
    \end{lstlisting}
    
  \item \xhdr{Visually Inspect the Dataframe} 
    
    We can plot the \emph{percentages} dataframe to see the differences better. 

    \begin{lstlisting}[caption={}, label={lst:pythoncode}]
import matplotlib.pyplot as plt
import numpy as np
# Get the skin ratings and values from the DataFrame
skin_ratings = percentages.index.values
values = percentages.values
# Set the width of the bars
bar_width = 0.2
# Set the x positions of the bars
r1_pos = np.arange(len(skin_ratings))
r2_pos = r1_pos + bar_width
r3_pos = r2_pos + bar_width
r4_pos = r3_pos + bar_width
# Create the bars for each value
plt.bar(r1_pos, values[:, 0], width=bar_width, label='Red Cards 1')
plt.bar(r2_pos, values[:, 1], width=bar_width, label='Rater 1')
plt.bar(r3_pos, values[:, 2], width=bar_width, label='Red Cards 2')
plt.bar(r4_pos, values[:, 3], width=bar_width, label='Rater 2')
# Set the title and axis labels
plt.title('Values by Skin Rating')
plt.xlabel('Skin Rating')
plt.ylabel('Value')
# Set the x-axis tick labels to the skin ratings
plt.xticks(r2_pos + bar_width / 2, skin_ratings)
# Set the legend
plt.legend()
# Show the plot
plt.show()
    \end{lstlisting}

    \item \xhdr{Controlling for LeagueCountry}

    To control for the \emph{leagueCountry} variable in the OLS model, you can include it as a categorical variable in the model. In Python, you can create dummy variables for the \emph{leagueCountry} variable using the \emph{get\_dummies()} function from pandas. Here's an example code to add \emph{leagueCountry} as a categorical variable:
    \begin{lstlisting}[caption={}, label={lst:rcode}]
import pandas as pd
import statsmodels.formula.api as smf
# Create dummy variables for leagueCountry
df = pd.get_dummies(df, columns=['leagueCountry'], drop_first=True)
# Fit a linear model
model = smf.ols("redCards ~ dark_skin_tone + age + leagueCountry_England + leagueCountry_Germany + leagueCountry_Spain", data=df).fit()
# Print the model summary
print(model.summary())    
\end{lstlisting}
In this example, \emph{leagueCountry} is added as three separate dummy variables for \emph{England}, \emph{Germany}, and \emph{Germany}, with \emph{Germany} being the reference category. The \emph{drop\_first=True} argument removes one of the dummy variables to avoid multicollinearity.

\end{itemize}

\section{Assistant Interface Implementation}
\label{appendix:implementation}
We implemented the wizard and user interface for the Jupyter environment \cite{Prez2007IPythonAS} as a JupyterLab extension. We extended the JupyterLab Commenting and Annotation extension \cite{jupyterlabcomments} and incorporated two modes: wizard mode and analyst mode. Both modes work with the same underlying notebook file and share a similar side panel interface. In \textit{wizard mode}, the wizard can highlight code or cells and add corresponding suggestions to the side panel. All suggestions were stored on disk as an underlying suggestion file. The extension, when in \textit{analyst mode}, polled this file for changes and updated the side panel when new suggestions were added to the file. 

In the study, both the analyst and wizard worked with the same notebook in the same file system.  We hosted the notebook on the wizard's machine. Accordingly, we connected the analyst’s machine remotely to the wizard’s machine using SSH. To ensure the wizard did not update actual notebook contents, we gave wizard mode no edit permissions.

\section{Notebook Details for Lab Study}
\label{appendix:notebook_detail}
The notebook included a  description of the task, the dataset columns, and associated descriptive statistics. 
% \tim{could consider putting this in appendix} 
In addition, the notebook had a profile report of the dataset which we created using the ydata-profiling library \cite{ydata-profiling}. This report included univariate analysis (i.e., descriptive statistics and distribution histograms) and multivariate analysis (i.e., correlations, missing data, duplicate rows) of the dataset columns. We wanted to help analysts quickly familiarize themselves with the dataset and direct their focus toward reasoning about their analysis approach.

\section{Study Observations for ``Where'' -- Location of Assistant and Suggestions}
\label{appendix:results_where}
While we did not explicitly ask analysts their preferences for the location of the assistant in the user interface, 7/13 analysts did express what they preferred. Most of these analysts mentioned that they liked the assistant being to the side of the notebook [A4, A6, A8, A11, A12]. For instance, A11 liked that the side panel allowed them to focus on the notebook: \shortquote{It's pretty useful that it is on the side and it doesn't (interfere) with what you are thinking what you are typing.} On the other hand, A3 and A10  preferred the suggestion closer to the cell.
% A3 found it a sub-optimal use of space: \shortquote{Because once I have read the suggestion, it's no longer useful. If it's taking the side of my screen, it's lying uselessly there.} 
% \tim{were they not aware of toggle/resizing functionality?} 

With respect to the location of the actual suggestions, analysts sometimes found it to be too cluttered [A5, A7, A9, A13]. This partially led to some analysts missing suggestions [A5, A10]. For A5, this became more difficult as it was \shortquote{very hard to view when you have a lot of things.} Thus, analysts suggested better ways of organizing the suggestions. For example, A13 wanted to tag variables with a categorization: \shortquote{It would be great to have a category of suggestions so I can quickly see if it's about processing variables or debugging or explanation.} A5 expressed similar views but would have liked \shortquote{some kind of color coding like how in Stack Overflow you have questions and answers in different colors.}

The lack of consensus on assistant location supports prior findings~\cite{McNutt2023OnTD}, in that it depends on the needs and preferences of the analyst.

%%%% 10-appendix.tex ends here %%%%

\end{document}